\documentclass[10pt,twocolumn]{article} 
\usepackage{simpleConference}
\usepackage{times}
\usepackage{graphicx}
\usepackage{booktabs}
\usepackage{amssymb}
\usepackage{tabularx}
\usepackage{url,hyperref}
\usepackage{amsmath,amsfonts}

\newcommand{\argminD}{\arg\!\min} 

\begin{document}

\title{Exploring the Power of Generative Deep Learning for Image-to-Image Translation and MRI Reconstruction: A Cross-Domain Review}

\author{Yuda Bi \\
\\
Center for Translational Research in Neuroimaging and Data Science (TReNDS) \\
Atlanta, Georgia, USA \\
\\
\\
ybi3@gsu.edu \\
}

\maketitle
\thispagestyle{empty}

\begin{abstract}
Deep learning has become a prominent computational modeling tool in the areas of computer vision and image processing in recent years. This research comprehensively analyzes the different deep-learning methods used for image-to-image translation and reconstruction in the natural and medical imaging domains. We examine the famous deep learning frameworks, such as convolutional neural networks and generative adversarial networks, and their variants, delving into the fundamental principles and difficulties of each. In the field of natural computer vision, we investigate the development and extension of various deep-learning generative models. In comparison, we investigate the possible applications of deep learning to generative medical imaging problems, including medical image translation, MRI reconstruction, and multi-contrast MRI synthesis. This thorough review provides scholars and practitioners in the areas of generative computer vision and medical imaging with useful insights for summarizing past works and getting insight into future research paths.
\end{abstract}

\section{Introduction}
The integration of deep learning techniques has enabled the application of convolutional neural networks (CNNs) for efficient computer vision. CNNs have proven to be effective in various computer vision tasks, including image classification \cite{AlexNet}, semantic segmentation \cite{long2015fully}, object detection \cite{ren2015faster} and medical image analysis \cite{U-Net}, \cite{li2014medical}. Another promising deep learning technique for generating new data through unsupervised and supervised learning of complex distributions is generative adversarial networks (GANs) \cite{GAN}. GANs consist of two neural networks, a generator, and a discriminator, that work in an adversarial manner. One of the key research directions for GANs is natural image-to-image (I2I) translation and synthesis, which is the process of converting one image into another while preserving or changing its complex and high-level semantic content. The purpose of I2I translation is to enable machines to comprehend and generate realistic images that are similar to the input images but have distinct visual qualities. I2I translation is an important AI-related direction that can reduce a large amount of human labor in certain works in daily life and the medical domain. As we already know, deep CNN models can extract high-level image features; therefore, GAN-based models combined with CNN in the generator can recover learned features from unprocessed data. I2I translation includes multiple directions in the domain of natural imaging, including style transfer \cite{luan2017deep},\cite{kim2017learning},\cite{yi2017dualgan}, image and video colorization \cite{zhang2016colorful},\cite{zhang2017real},\cite{suarez2017infrared}, image Super-Resolution \cite{zhu2020gan},\cite{yuan2018unsupervised}, etc. GANs have been increasingly applied to medical image translation tasks \cite{uzunova2020memory} and have produced results comparable to conventional I2I translations. They are gradually replacing traditional non-generative deep learning approaches in various medical image analysis tasks involving different modalities, such as magnetic resonance imaging (MRI), computerized tomography (CT), and X-ray \cite{kspace}.

MRI is a cutting-edge medical imaging technique that has revolutionized medicine, radiology, and psychiatry \cite{kspace}. With its ability to provide quantifiable and reproducible information about the structural, anatomical, and functional properties of tissue, MRI has surpassed CT scans and X-rays in terms of image clarity and patient safety, as it produces clearer images of organs and soft tissues without exposing the patient to radiation. However, the high complexity of the MRI imaging process can result in prolonged acquisition times due to the long process of generating high-resolution MR images from fully-sampled k-space. In contrast to conventional medical I2I translation, MRI reconstruction seeks to overcome this issue by creating de-aliased MR images using under-sampled k-space rather than fully-sampled k-space. This approach preserves the original signal for greater clinical usage while reducing the acquisition duration of an MR image. Compressed sensing (CS), a mathematical theory used for signal acquisition and processing, has been commonly used for this purpose \cite{compressedsensing}. Although reconstructing full-sampled images from under-sampled k-space is an ill-posed linear inverse task, incorporating assumptions or prior information, such as the sparsity of the frequency domain (k-space) of the MR image, can help regularize the solution by formulating a convex optimization problem. Methods such as SENSE \cite{sense} and GRAPPA \cite{grappa}, based on CS-MRI, have achieved remarkable success in the past decades. 

As deep learning techniques gained popularity in the computer vision area, researchers sought to apply these methods to MRI reconstruction instead of relying on traditional CS-MRI methods. CNNs were able to learn a mapping from under-sampled k-space to the image domain, which enabled them to reconstruct the original MR image. Despite the automatic learning process in CNNs, the significant manual effort was still required to design effective losses. For instance, in some generative tasks, CNNs faced challenges in achieving good performance due to the difficulty in defining a suitable loss function. In contrast, GANs can be used to specify only a high-level goal, such as "make the output indistinguishable from reality," and then automatically learn an appropriate loss function to meet that goal \cite{cGANimagetoimage}. As with other areas, researchers in the field of MRI reconstruction began utilizing GANs instead of CNNs, leading to substantial advancements. Nevertheless, medical imaging still presents numerous difficulties compared to natural imaging-based tasks. One major challenge is the scarcity of large datasets like ImageNet \cite{russakovsky2015imagenet} for medical imaging, which is largely due to data privacy restrictions and protection policies. From a technical standpoint, medical imaging has physiological and anatomical significance, making the learning process more complex and requiring additional interpretation. This adds an extra layer of complexity to the field of medical imaging, compared to natural imaging.

The purpose of our survey is to provide a comparative overview of GAN-based models in I2I translation and MRI reconstruction. In Section II, we present a background on GANs, I2I Translation, and CS-MRI reconstruction. Section III reviews the literature on I2I Translation using deep learning models and provides in-depth information about Pix2Pix. Section IV focuses on the most well-known and benchmarked GAN-based deep learning models for MRI reconstruction. Section V explores more complex medical imaging reconstruction problems, such as multi-contrast MRI reconstruction. In Section VI, we outline commonly used evaluation metrics in I2I translation and MRI reconstruction and provide important considerations for conducting experiments in these fields. Finally, in Section VII, we discuss the challenges in I2I translation and MRI reconstruction and suggest potential future directions for research.

\section{Background}

\subsection{Generative Adversarial Network}
The Generative Adversarial Network (GAN) \cite{GAN} has proven to be effective in several applications, such as I2I translation, image super-resolution, image reconstruction, and data augmentation. GANs are inspired by two-player zero-sum games in game theory and consist of two components: the generator $G$ and the discriminator $D$. The generator $G$ is designed to learn the mapping from a noise distribution $p_z(z)$ to the data space $p_{data}$, while the discriminator $D$ operates as a binary classifier that distinguishes between the generated fake sample $G(z)$ from the distribution $p_g$ and the real sample $\mathbf{x}$ from the distribution $p_{data}$. The loss function of the GAN model can be expressed as:
\begin{align}
    \min _G \max _D L(G,D) = \mathbb{E}_{x\sim p_{data}(\boldsymbol{x})}[\log D(\boldsymbol{x})] \\ \nonumber
    + \mathbb{E}_{z\sim p_{z}(\boldsymbol{z})}[\log(1-D(G(\boldsymbol{z}))]
\end{align}
Where ${E}_{x\sim p_{data}}$ is the expected value over all sampled real data instances, ${E}_{z\sim p_{z}(\mathbf{z})}$ is the expected value over all generated fake instances $G(z)$.
Training a GAN model involves simultaneously training the generator $G$ and discriminator $D$. Given a batch of noise samples ${z^{(1)},...,z^{(n)}}$ from the noise distribution $p_z(z)$ and a batch of data samples ${x^{(1)},...,x^{(n)}}$ from the data space $p_{data}$, the discriminator $D$ is updated using its stochastic gradient ascent:
\begin{equation}
\nabla_{\mathbf{W}_D} \frac{1}{n} \sum_{i=1}^n\left[\log D\left(\boldsymbol{x}^{(i)}\right)+\log \left(1-D\left(G\left(\boldsymbol{z}^{(i)}\right)\right)\right)\right]
\end{equation}
After updating the discriminator $D$, we need to update the generator $G$ by:
\begin{equation}
\nabla_{\mathbf{W}_G} \frac{1}{n} \sum_{i=1}^n \log \left(1-D\left(G\left(\boldsymbol{z}^{(i)}\right)\right)\right)
\end{equation}
Where $\mathbf{W}_D$ and $\mathbf{W}_G$ represents the trainable parameters from discriminator $D$ and generator $G$. 

\subsection{Conditional Generative Adversarial Network}

The Conditional Generative Adversarial Network (CGAN) \cite{cGAN} is a GAN variation incorporating a conditional element. In a CGAN, the generator $G$ and discriminator $D$ are conditioned on additional input data, such as an image or a label. This allows the CGAN to generate samples that are conditioned on specific characteristics, such as specific object classes in an image or specific attributes in a dataset. The objective function of a CGAN can be written as follows: 
\begin{align}
    \min _G \max _D V(D, G)=\mathbb{E}_{\boldsymbol{x} \sim p_{\text {data }}(\boldsymbol{x})}[\log D(\boldsymbol{x} \mid  \boldsymbol{y})]+ \\ \nonumber
    \mathbb{E}_{\boldsymbol{z} \sim p_z(\boldsymbol{z})}[\log (1-D(G(\boldsymbol{z} \mid \boldsymbol{y})))]
\end{align}

The condition $\boldsymbol{y}$ could be a label or an image, $z$ is the random noise, and $x$ represents the ground-truth image. The generator $G$ takes the condition $\boldsymbol{y}$ and the random noise $z$ as input, while the discriminator $D$ receives the data pairs ($\boldsymbol{y}$, $G(\boldsymbol{y}$, $z$)) and ($\boldsymbol{y}$, $x$) and then determines which is real and which is fake. Figure 1 illustrates the architecture of both GANs and CGANs.

\begin{figure}[ht]
\centering
  \includegraphics[scale=0.45]{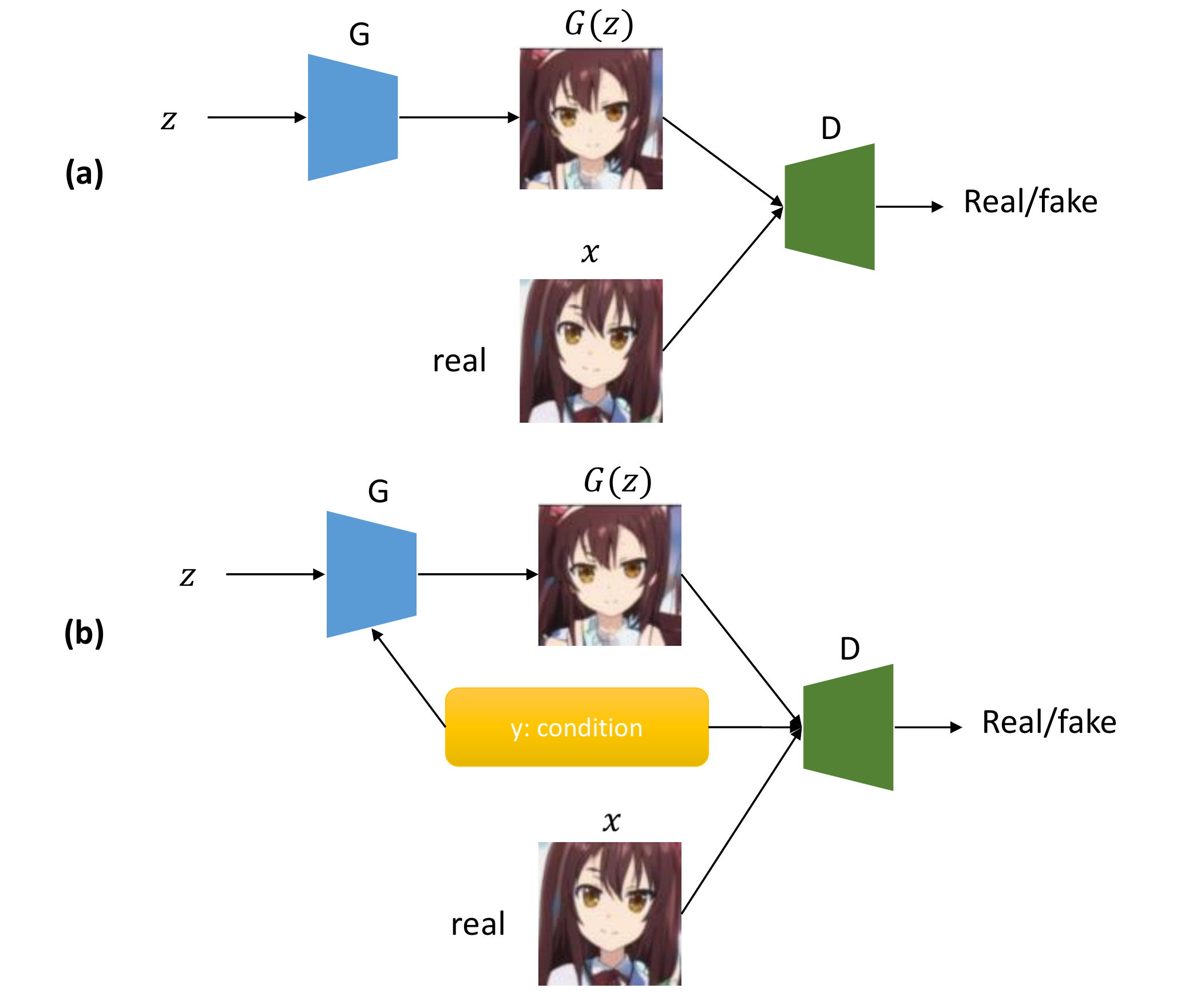}
  \caption{(a) GAN architecture (b) Conditional GAN (CGAN) architecture}
\end{figure}

\subsection{CS-MRI reconstruction}

\begin{figure*}[ht]
\centering
  \includegraphics[scale=0.4]{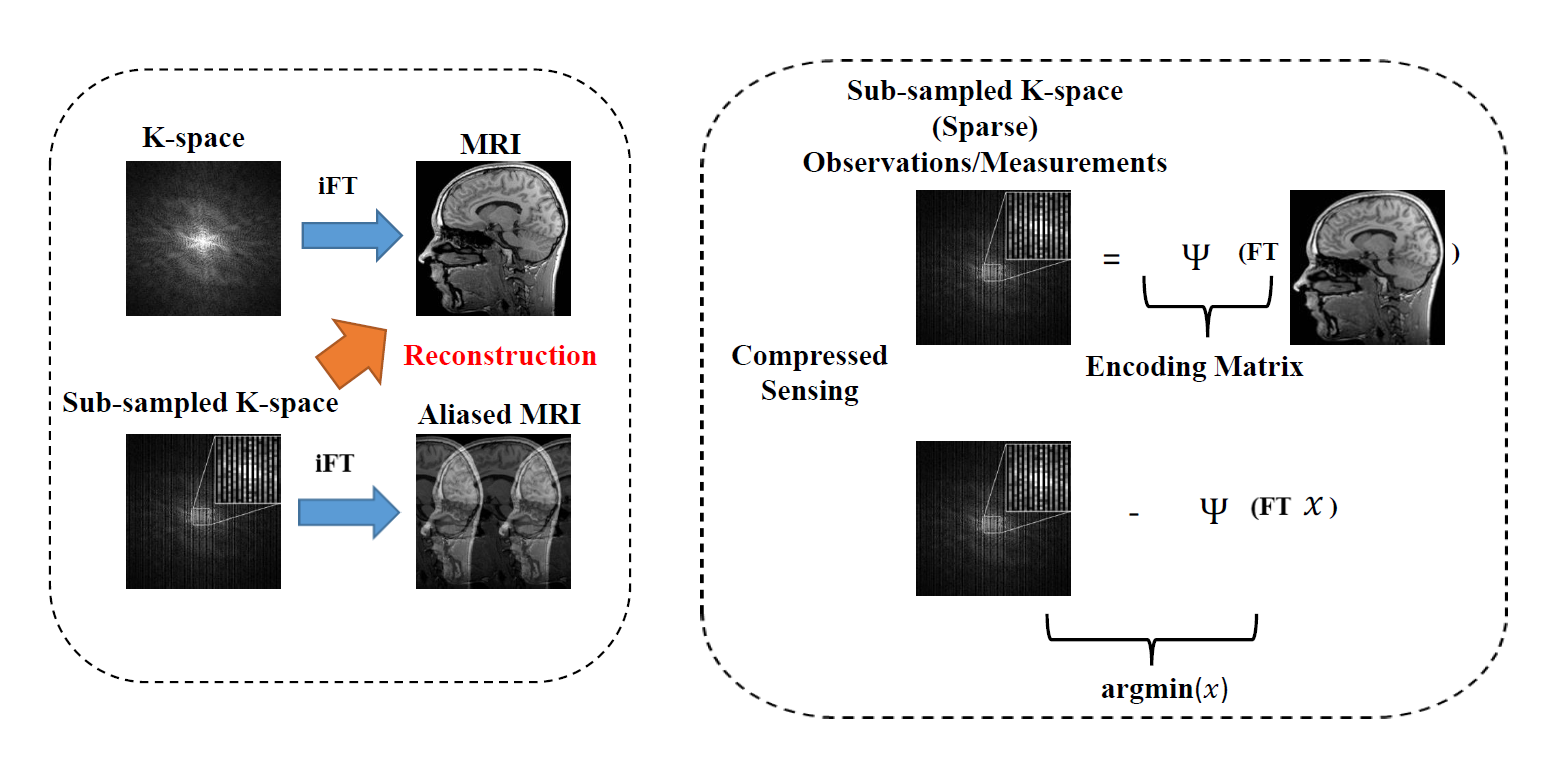}
  \caption{The task of MRI reconstruction(left) and the general problem description of compressed sensing(right).}
\end{figure*}

CS-MRI (Compressed Sensing MRI) reconstruction is a method for acquiring MRI images with reduced data acquisition time, which can be particularly useful in applications where prolonged scanning times are not feasible, such as with patients who have difficulty remaining still for extended periods of time \cite{jaspan2015compressed}. CS-MRI operates by sparsely sampling the k-space data, the frequency domain representation of the MRI image, i.e., acquiring a small number of samples. This produces an under-sampled k-space, which can then be reconstructed using advanced mathematical algorithms, such as sparse optimization algorithms or deep learning-based methods, to produce an image. Certain image features can be reconstructed from a sparse set of k-space samples, reducing the amount of data needed to obtain an acceptable image quality. This can lead to faster scan times, enhanced patient comfort, and cost savings.

Given a MR image $\mathbf{x}\in\mathbb{C}^{N_1 \times N_2}$, we do a fully-sampled measurement $\mathbf{y}$ on k-space, using the Fourier transform $\mathcal{F}(\cdot)$:
\begin{equation}
    \mathbf{\hat{y}} = \mathcal{F}\mathbf{x} + \sigma
\end{equation}
Where $\sigma$ represents noise. Using compressed sensing, we can reconstruct $\mathbf{x}$ from undersampled k-space measurement $\mathbf{\hat{y}}$ due to the sparsity characteristic of the MR image in its frequency domain. So that:
\begin{equation}
    \mathbf{\hat{y}} = \mathcal{F}\mathbf{x} + \sigma
\end{equation}
if we use a binary sampling mask $\mathcal{M}\in\mathbb{C}^{N_1 \times N_2}$ is a method for selecting a subset of k-space lines in order to accelerate the measurement, then:
\begin{equation}
    \mathbf{\hat{y}} = \mathcal{M}\times \mathcal{F}\mathbf{x} + \sigma
\end{equation}
The sampling mask and the Fourier transform can be combined as an encoding matrix $\Phi$, in order to reconstruct $\mathbf{x}$, we need to minimize the noise, so that:
\begin{equation}
    \mathbf{\hat{x}} =  \argminD_{\mathbf{x}} \|\Phi\mathbf{x} - \mathbf{\hat{y}}\|^2_2
\end{equation}
Instead of using compressed sensing, non-generative DL-based approaches, such as CNNs, can be formulated as an optimization problem:
\begin{equation}
    \mathbf{\hat{x}} =  \argminD_{\mathbf{x}} \|\Phi\mathbf{x} - \mathbf{\hat{y}}\|^2_2 + \lambda\|\mathbf{x}-f_{DL}(\mathbf{x}_0|\phi)\|^2_2
\end{equation}
Where $\mathbf{x}_0$ represents the zero-filled reconstruction, $f_{DL}(\mathbf{x}_0|\phi)$ is the output image from the trained model, $\phi$ is the optimized parameters in the trained model. Figure 2 shows the description of CS-based MRI reconstruction. 

However, CS-MRI reconstruction has two major limitations: it requires a significant amount of computational time, and it is heavily dependent on the choice of regularization parameters and transforms basis, which must be adjusted carefully \cite{lv2021gan}.

\begin{figure*}[t]
\centering
  \includegraphics[scale=0.6]{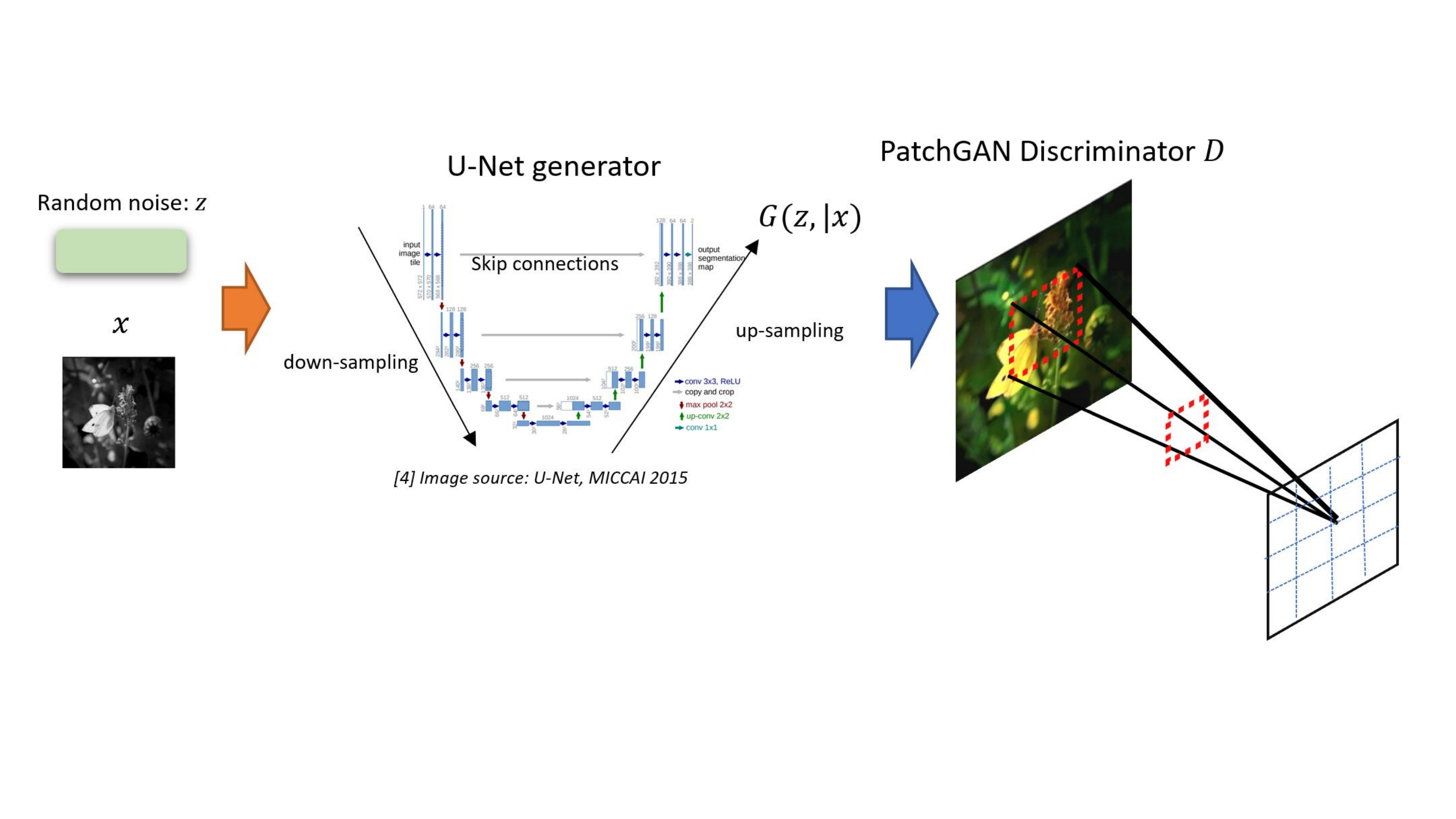}
  \caption{cGAN-based architecture for Pix2Pix}
\end{figure*}

\section{Image-to-Image Translation with Deep Learning}

I2I translation is a subfield of computer vision and deep learning that seeks to convert images from one domain to another while maintaining their content and structure. Numerous tasks, including style transfer, semantic segmentation, object detection, image synthesis, and data augmentation, have found extensive applications for this technology. The continuous development of I2I translation promises to uncover additional potential applications in a variety of industries, such as the entertainment industry, healthcare, and autonomous vehicles. This section begins with a review of the literature on I2I translation using deep learning techniques. It also introduces two well-known GANs, Pix2Pix and CycleGAN, which are primarily concerned with image style transfer.

\subsection{Related Works:} Previous works have shown the effectiveness of CGAN for I2I translation tasks. One notable example is the Pix2Pix model \cite{cGANimagetoimage}, which utilizes a CGAN to produce remarkable results in various applications, such as transforming satellite views into street maps and generating shoes from border sketches. The generator in Pix2Pix has a U-Net-like architecture that allows it to perform style transfer based on paired training images while preserving the pixel-wise characteristics of the source image. This results in accurate I2I translations combining high-level semantic and detailed pixel information. Zhu et al. \cite{zhu2017unpaired} introduced CycleGAN, which used unpaired images as training samples combined with a cycle-consistency loss, producing more excellent results on image art style transfer. Wang et al. \cite{wang2018high} proposed a new CGAN, HDPix2Pix, for high-resolution image synthesis, which can generate photo-realistic images without the need for hand-crafted losses or pre-trained networks. Almahairi et al. \cite{almahairi2018augmented} then developed Augmental CycleGAN, which enables many-to-many mapping using unpaired data. Huang et al. \cite{huang2018multimodal} proposed MUNIT, a framework based on unsupervised I2I translation. MUNIT has outperformed CycleGAN in various I2I tasks. However, these related works to Pix2Pix have limitations in scalability and robustness when dealing with multiple domains. To address this, Choi et al. \cite{choi2018stargan} proposed StarGAN, which can perform I2I translations for multiple domains using a single model. A 'domain' is defined as images sharing the same attribute value, such as images of women or people with black hair. StarGAN v2 \cite{choi2020stargan} is an updated version of the original StarGAN model \cite{choi2018stargan}. StarGAN v2 uses the Adaptive Layer-Instance Normalization (AdaLIN) generator architecture, allowing the model to adjust the normalization parameters for each instance, improving performance for diverse domains. Additionally, StarGAN v2 has the capability to perform I2I translations in an unsupervised manner using unpaired training data. More recently, Cao et al. \cite{cao2021remix} introduced the Remix framework, which overcomes the over-fitting problem encountered with fewer training examples/paired images. The ReMix method can easily be incorporated into existing GAN models with minimal modifications. Table I shows some important GAN architectures for I2I translation. 

\begin{table*}[]
\caption{Important GAN models for I2I translation}
\centering
\begin{tabular}{@{}lllll@{}}
\toprule
Reference                                 & Year & Dataset                                        & Domain      & highlight                                               \\ \midrule
\multicolumn{1}{l|}{Pix2Pix \cite{cGANimagetoimage}}           & 2017 & Paired data                  & Two        & First well-known GAN for I2I translation\\
\multicolumn{1}{l|}{CycleGAN \cite{zhu2017unpaired}}        & 2017 & unpaired data                                          & Two    & unpaired domain transfer  \\
\multicolumn{1}{l|}{Dual-GAN \cite{yi2017dualgan}}        & 2017 & unpaired data                                          & Two    & unpaired domain transfer  \\
\multicolumn{1}{l|}{HDpix2pix \cite{wang2018high}}           & 2018 & paired data             &Two    & high-resolution improved Pix2Pix             \\
\multicolumn{1}{l|}{Attention GAN \cite{chen2018attention}}            & 2018 & unpaired data & Multiple        & Attention-based                              \\ 
\multicolumn{1}{l|}{StarGAN \cite{choi2018stargan}}            & 2018 & unpaired data & Multiple        & Multiple domains translation                  \\ 
\multicolumn{1}{l|}{TravelGAN \cite{amodio2019travelgan}}            & 2019 & unpaired data & Two        & Vector learning instead of cycle consistency                  \\ 
\multicolumn{1}{l|}{SelectionGAN \cite{tang2019multi}}            & 2019 & paired data & Two        & Cross-view image translation                   \\ 
\multicolumn{1}{l|}{StarGAN v2 \cite{choi2020stargan}}            & 2020 & unpaired data & Multiple        & Multiple domains translation                             \\ 
\multicolumn{1}{l|}{SPA-GAN \cite{emami2020spa}}            & 2020 & unpaired data & Multiple        & Spatial attention                             \\ 
\bottomrule
\end{tabular}

\end{table*}

\subsection{Pix2Pix}

Pix2Pix is a pioneering and groundbreaking I2I translation infrastructure. It is a large-scale, systematic I2I project that implemented supervised paired data training for two-domain image translation using an image-based conditional GAN. Pix2Pix employs U-net as the generator backbone to preserve low-level and high-level structural and textural information and PatchGAN as the discriminator capable of handling large images based on the fixed-size patch discriminator. Pix2Pix has been trained on a number of paired image datasets, including the Cityscapes dataset for semantic label vs. photo, the CMP Facades dataset for architectural label vs. photo, map vs. aerial photo from Google Map, and sketch and edge vs. photo. These models performed well during validation and testing and also provide an online resource for anyone to conduct their own trials.

\textbf{General definitions: }Pix2Pix uses an image-CGAN architecture that consists of two networks: a generator $G$ and a discriminator $D$. The aim of Pix2Pix is to translate an input image to a target image by learning a mapping and distribution based on a condition image. The generator network takes a conditional input image $\boldsymbol{x}$ and a random noise $\boldsymbol{z}$ as input. It generates a corresponding output image $\boldsymbol{\hat{y}}$, while the discriminator network assesses the realism of the generated image compared to the ground-truth image $\boldsymbol{y}$. Pix2Pix has been applied to various I2I translation tasks, such as converting maps to satellite images, sketches to photos, and gray-scale images to color images. Figure 3 demonstrates the details of the Pix2Pix architecture with an example of converting sketches to real objects. The objective and loss function of Pix2Pix is similar to those of a CGAN, which can be written as:
\begin{equation}
\begin{aligned}
\mathcal{L}_{c G A N}(G, D)= & \mathbb{E}_{x, y}[\log D(x
, y)]+ \\
& \mathbb{E}_{x, z}[\log (1-D(x, G(x, z))],
\end{aligned}
\end{equation}
Additionally, as noted in previous research \cite{pathak2016context}, the generator can sometimes produce images that are too dissimilar from the target images, making it easy for the discriminator to identify them as fake. To address this issue, including an $L_1$ or $L_2$ loss term in the generator's loss function is common. This loss term measures the difference between the generated image and the target image and penalizes the generator for producing images that are significantly different from the target. As such, an $L_1$ or $L_2$ loss term should be added to the generator $G$ itself. In Pix2Pix, $L_1$ loss is used to prevent the blurring of images: 
\begin{equation}
\mathcal{L}_{L 1}(G)=\mathbb{E}_{x, y, z}\left[\|y-G(x, z)\|_1\right]
\end{equation}
So that the final loss function to optimize the model is: 
\begin{equation}
G^*=\arg \min _G \max _D \mathcal{L}_{c G A N}(G, D)+\lambda \mathcal{L}_{L 1}(G)
\end{equation}

\textbf{U-Net generator: }The Pix2Pix generator uses the U-Net architecture \cite{U-Net}, which was state-of-the-art for image segmentation at the time of its development. The U-Net architecture consists of a series of down-sampling and up-sampling convolutional layers. The use of the U-Net architecture in the generator provides several benefits. One key advantage is its ability to handle large amounts of spatial information, making it well-suited for medical image segmentation tasks where the input images can be very large. Additionally, the symmetrical architecture of the U-Net enables the efficient transfer of information from the down-sampling path to the up-sampling path, preserving fine details of the segmentation mask. The U-Net also includes skip connections, which are implemented by concatenating activation from an earlier layer to the activation of a deeper layer. This allows the network to access high-level and low-level features, providing a complete representation of the input image. The skip connections also help capture long-range dependencies in the data by combining features from the contracting path with those from the expanding path.

\textbf{PatchGAN discriminator:} PatchGAN is a type of GAN discriminator that penalizes local image patch structures. It categorizes each $N \times N$ patch in an image as either real or fake. The size of the patch can be adjusted as needed. One benefit of the PatchGAN discriminator is its ability to apply a fixed-size patch discriminator on images of any size, unlike typical discriminators that only score the full image. Additionally, a smaller PatchGAN has fewer parameters, runs faster, and can be applied to images of any size.

\textbf{Training and optimization: } GAN models are notoriously challenging to train and converge due to the simultaneous training of the generator and discriminator, with improvements to one model often coming at the expense of the other. The differing learning rates of the two models can also make convergence difficult. As suggested in the original GAN paper \cite{GAN}, instead of training the generator $G$ to minimize $\log(1-D(x,G(x,z)))$, it is more effective to train $G$ to maximize $\log(D(x,G(x,z)))$ to improve convergence. In Pix2Pix, using instance normalization, which involves setting the batch size to 1, is a better choice than batch normalization for training generative models \cite{ulyanov2016instance}.

\subsection{CycleGAN}

\textbf{General definition: }CycleGAN \cite{zhu2017unpaired} is another well-known I2I translation benchmark. It extends Pix2Pix by enabling unpaired data domain transmission, thereby resolving the issue of insufficient data. CycleGAN's objective, given two distinct image domains $X$ and $Y$, is to learn the mapping between the two domains so that an image from domain $X$ can be transformed into an image in domain $Y$ while preserving the original image's content, and vice versa. It is used to enforce the cyclic consistency constraint between two image domains. If we translate an image from the domain $X$ to the domain $Y$ and then back to the domain $X$, we should obtain the original image. This constraint ensures that the generated images contain meaningful information and are not merely arbitrary noise. Figure 4 depicts the architecture of CycleGAN and its Cycle-consistency loss. 
\begin{figure*}[t]
\centering
  \includegraphics[scale=0.7]{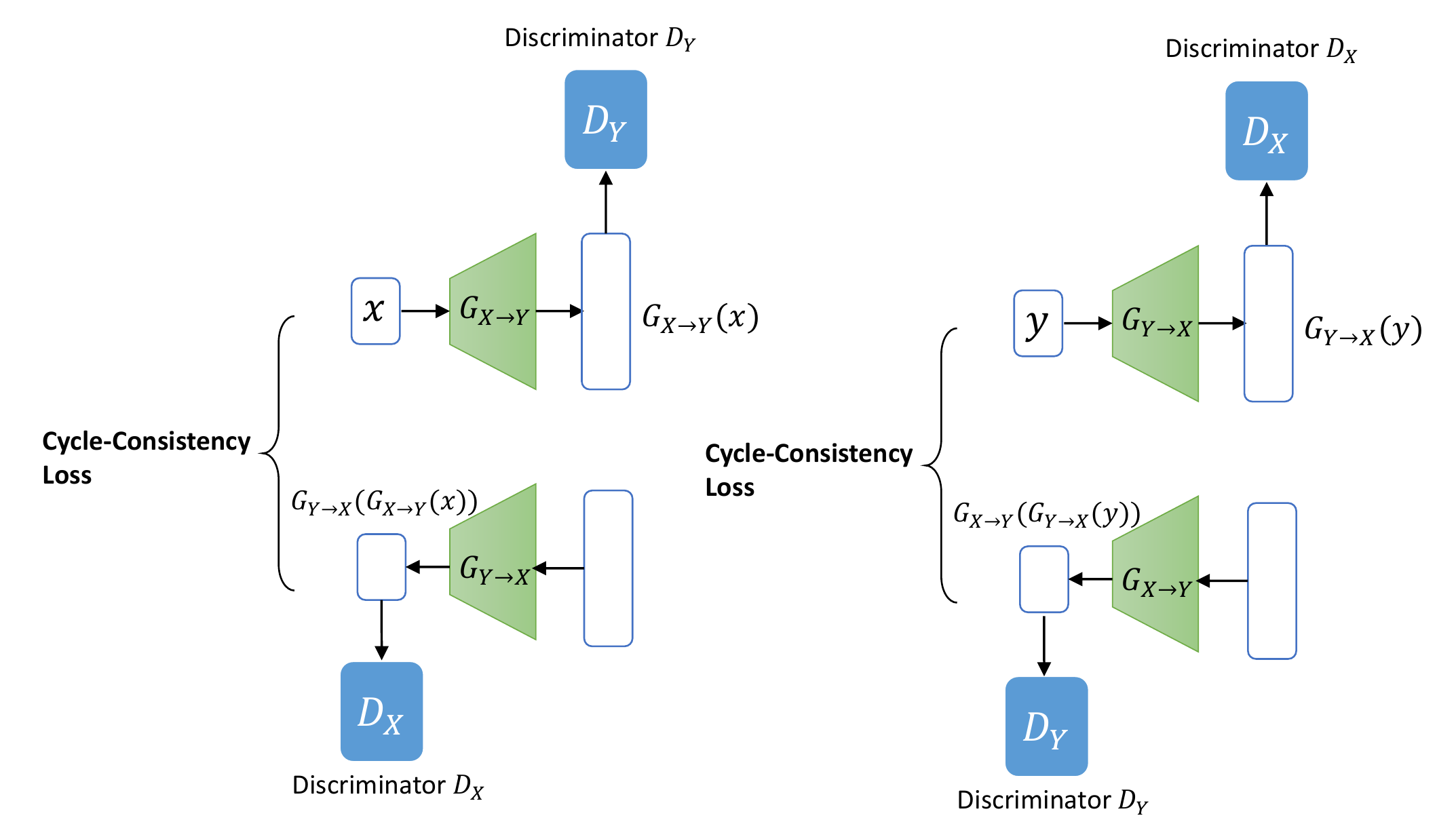}
  \caption{GAN architecture and Cycle-consistency loss of CycleGAN. }
\end{figure*}

\textbf{Loss function:} There are two GANs in CycleGAN pipeline: $G_{X \rightarrow Y}$ represents the generator that map the image from domain $X$ to $Y$ and the discriminator $D_Y$ tries to identify the fake image $G_{X \rightarrow Y}(x)$ and image $y$ from domain $Y$; $G_{Y \rightarrow X}$ represents the generator that map the image from domain $Y$ to $X$ and the discriminator $D_X$ tries to identify the fake image $G_{Y \rightarrow X}(y)$ and image $x$ from domain $X$. So there are two adversarial losses; for example, from the generator: 

\begin{equation}
\begin{aligned}
\mathcal{L}_{\mathrm{GAN}}\left(G_{XY}, D_Y, X, Y\right) & =\mathbb{E}_{y \sim p_{\text {data }}(y)}\left[\log D_Y(y)\right] \\
& +\mathbb{E}_{x \sim p_{\text {data }}(x)}\left[\log \left(1-D_Y(G_{XY}(x))\right]\right.
\end{aligned}
\end{equation}

Meanwhile, the GAN that maps the image in domain $Y$ to $X$ also has an objective function:

\begin{equation}
\begin{aligned}
    \mathcal{L}_{\mathrm{GAN}}\left(G_{YX}, D_X, X, Y\right) & =\mathbb{E}_{y \sim p_{\text {data }}(x)}\left[\log D_Y(y)\right] \\
& +\mathbb{E}_{x \sim p_{\text {data }}(x)}\left[\log \left(1-D_X(G_{YX}(y))\right]\right.
\end{aligned}
\end{equation}

However, these two losses do not guarantee that the translated image will retain the domain-specific information. For instance, an image from domain $X$ that has been translated to domain $Y$ may share similar characteristics with images in the $Y$ domain while discarding all information from domain $X$. CycleGAN uses a specific loss function to let the networks keep domain information, which is called cycle-consistency loss. The idea of forwarded cycle-consistency loss is that for each image $x$ from domain $X$, the image translation cycle should be able to bring $x$ back to the original image, which means to minimize the distance between $x$ and $G_{Y \rightarrow X}(G_{X \rightarrow Y}(x))$. Similarly, the backward cycle-consistency loss is to minimize the distance between $y$ and $G_{X \rightarrow Y}(G_{Y \rightarrow X}(y))$. The cycle-consistency loss can be expressed by:

\begin{equation}
\begin{aligned}
\mathcal{L}_{\text {cyc }}(G_{XY}, G_{YX}) & =\mathbb{E}_{x \sim p_{\text {data }}(x)}\left[\|G_{YX}(G_{XY}(x))-x\|_1\right] \\
& +\mathbb{E}_{y \sim p_{\text {data }}(y)}\left[\|G_{XY}(G_{YX}(y))-y\|_1\right]
\end{aligned}
\end{equation}

So that the total objective can be represented by:
\begin{equation}
\begin{aligned}
\mathcal{L}\left(G_{XY}, G_{YX}, D_X, D_Y\right) & =\mathcal{L}_{\mathrm{GAN}}\left(G_{XY}, D_Y, X, Y\right) \\
& +\mathcal{L}_{\mathrm{GAN}}\left(G_{YX}, D_X, Y, X\right) \\
& +\lambda \mathcal{L}_{\mathrm{cyc}}(G_{XY}, G_{YX})
\end{aligned}
\end{equation}

CycleGAN is the first GAN-based unpaired I2I translation model, and it has made significant advances and has significant implications for future research. Importantly, the cycle-consistency loss has been widely used in future works \cite{liu2017unsupervised, wu2019transgaga, kim2019u, gokaslan2018improving}.    

\section{MRI Reconstruction with Deep Learning}

As stated in the prior chapter, I2I Translation has been utilized extensively in natural imaging. I2I also applies to medical imaging, including MRI and CT cross-domain translation. In contrast to I2I translation, MRI and multi-contrast MRI reconstruction attempts to recreate MR images using under-sampled k-space data rather than directly from other modalities.

\begin{table*}[]
\centering
\caption{Important selected Deep learning based models for MR image reconstruction}
\resizebox{\width}{!}{
\begin{tabular}{@{}llll@{}}
\toprule
Reference                                 & Year & Dataset                                        & Model                                                  \\ \midrule
\multicolumn{1}{l|}{Non-generative model} &      &                                                &                                                                     \\ \cmidrule(r){1-1}
\multicolumn{1}{l|}{Wang et al\cite{CNN4MRI1}}           & 2016 & 500 fully sampled brain MRI                  & CNN         \\
\multicolumn{1}{l|}{Palangi et al\cite{distributedlstm}}        & 2016 & MNIST                                          & LSTM+CS       \\
\multicolumn{1}{l|}{Yang et al\cite{yangCNN4MRI}}           & 2018 & 100 fully sampled brain MR images              & CNN+CS                \\
\multicolumn{1}{l|}{Schlemper\cite{cascadecnn}}            & 2017 & Fully sampled short-axis cardiac cine MR scans & CNN                          \\ \cmidrule(r){1-1}
\multicolumn{1}{l|}{GAN-based model}      &      &                                                &                                                                      \\ \midrule
\multicolumn{1}{l|}{Yang et al\cite{yang2017dagan}}           & 2017 & MICCAI 2013 grand challenge dataset            & VGG-GAN                \\
\multicolumn{1}{l|}{Mardani et al\cite{DGANforCS}}        & 2019 & Knee dataset                                                   \\
\multicolumn{1}{l|}{Quan et al\cite{cyclicloss}}           & 2018 & MRI                                            & GAN                                       \\
\multicolumn{1}{l|}{Jiang et al\cite{jiang2019accelerating}}           & 2019 & MRI                                            & WGAN                                        \\
\multicolumn{1}{l|}{Li et al\cite{SEGAN}}             & 2019 & MICCAI 2013 grand challenge dataset            & U-Net-GAN                \\
\multicolumn{1}{l|}{\textit{Murugesan} et al.\cite{murugesan2019recon}}            & 2019 & MRIs                            & Recon-GLGAN              \\
\multicolumn{1}{l|}{Shaul et al\cite{shaul2020subsampled}}          & 2020 & Multimodal MRI data                            & U-Net-GAN               \\
\multicolumn{1}{l|}{Chen et al\cite{chen2021wavelet}}          & 2021 & Multimodal MRI data               &    WPD-DAGAN         \\
\multicolumn{1}{l|}{Korkmaz et al\cite{SVtrans}}        & 2021 & T1-weighted brain MRI                          & ViT-GAN                   \\
\multicolumn{1}{l|}{Lin et al\cite{vitganfaster}}            & 2022 & knee and brain MRIs                            & ViT-GAN                           \\
 \midrule
\multicolumn{1}{l|}{Diffusion model}      &      &                                                &                                                                     \\ \cmidrule(r){1-1}
\multicolumn{1}{l|}{Cao et al\cite{diffusion1}}            & 2022 & fastMRI dataset                                               \\
\multicolumn{1}{l|}{Xie et al\cite{xie2022measurement}}                               & 2022 & fastMRI single-coil knee data                   & MC-DDPM                                       \\ \bottomrule
\end{tabular}}
\end{table*}

\subsection{Related Works}
CNN is a potent DL model that has been widely applied to various computer vision problems ever since Alex et al. developed one of the most well-known CNN architectures, AlexNet \cite{AlexNet}. CNN has evolved into a bench of permutations, such as VGGNet \cite{VGG}, ResNet \cite{ResNet}, U-Net \cite{U-Net}, etc. With the advent of GPUs, fairly deep CNNs such as ResNet have achieved remarkable success and can outperform most conventional techniques. However, CNNs are still considered black boxes at that time since they lack interpretation. Wang et al. \cite{CNN4MRI1} started to use CNN models instead of CS-MRI on the acceleration of MRI images by reducing the scanning time; they provide a supervised learning CNN model that can map the under-sampled images to fully-sampled images. Yang et al. \cite{yangCNN4MRI} proposed a model-driven novel CNN model called ADMM-CSNet by reformulating the ADMM algorithms; it achieved better results than some data-driven CNN models. 

After the CNN models become popular in the DL field, GANs can outperform conventional CNN models because they can learn data distributions significantly faster than CNN models. DAGAN \cite{yang2017dagan} is the first well-known supervised GAN framework with conditions for MRI reconstruction. It employs a U-Net-like \cite{U-Net} architecture with the skip connection in the generator, a refined learning algorithm for fast convergence, and a novel combined loss function for superior reconstruction with preserved perceptual image details. Mardani et al. \cite{DGANforCS} provided a deep GAN to combine with the CS model, which can reduce the high-frequency noise and map zero-filled MR images to high-resolution MR images. Quan et al. \cite{cyclicloss} designed a dual-bench generator with ReconGAN and RefineGAN combining cyclic data consistency loss and generative adversarial loss, which can correctly map
the under-sampled data to fully-reconstruction data. Li et al. \cite{SEGAN} proposed the Structure-Enhanced GAN (SEGAN) with a new structure regularization called patch correlation regularization that aims to restore structure information locally and globally for efficient extraction of structure information. Murugesan et al. \cite{murugesan2019recon} proposed a novel GAN framework called reconstruction global-Local GAN (Recon-GLGAN), containing a generator and a context discriminator which incorporates global and local contextual information from images. Shaul et al. \cite{GANforBrainMRI} introduced a practical, software-only GAN framework to estimate the missing k-space sample for brain MRI accelerating acquisition. 

Most of the GAN models we discussed above use CNN-based generators and discriminators, especially U-Net and ResNet. They usually improve performance by redesigning loss functions or regularization. Recently, with the development of Vision Transformer (ViT), widely used in computer vision tasks that can outperform traditional CNN architectures, Some researchers have begun using ViTs as backbones instead of CNNs in GAN models. Korkmaz et al. \cite{SVtrans} proposed a generative ViT called GVTrans, which can progressively map low-dimensional noise and latent variables onto MR images via cascaded blocks of cross-attention ViTs. Zhao et al. \cite{zhao2023swingan} demonstrated a GAN model called SwinGAN, which uses a Swin-Transformer-based backbone as the generator. Swin Transformer \cite{swintransformer} is a U-Net shape hierarchical architecture ViT, which achieved robust results on various downstream tasks. Importantly, Lin et al. \cite{vitganfaster} given potential proof of the ViT-based GAN model can enable robust and fast accelerated MRI, using a pre-trained ViT model on natural images datasets, such as ImageNet, as the backbone of the GAN model that can be fine-tuned in MRI datasets which is costly to obtain. Table II shows some important GAN architectures for MRI reconstruction.

\subsection{DAGAN}

DAGAN \cite{DGANforCS} is one of the first de-aliasing and fast CS-MRI architectures that outperforms conventional CS-based methods and non-generative DL-based algorithms in MRI de-aliasing and reconstruction tasks. The DAGAN generator uses a CNN architecture similar to the U-Net \cite{U-Net} with skip connections, which is a highly advanced encoder-decoder network for medical image segmentation. DAGAN combines an adversarial loss with a novel content loss that considers both pixel-wise MSE and perceptual loss as defined by pre-trained VGGNet \cite{VGG}. Additionally, frequency domain information from the CS-MRI is used as an additional data consistency constraint. Figure 5 illustrates the architecture of DAGAN.

\begin{figure}[ht]
\centering
  \includegraphics[scale=0.6]{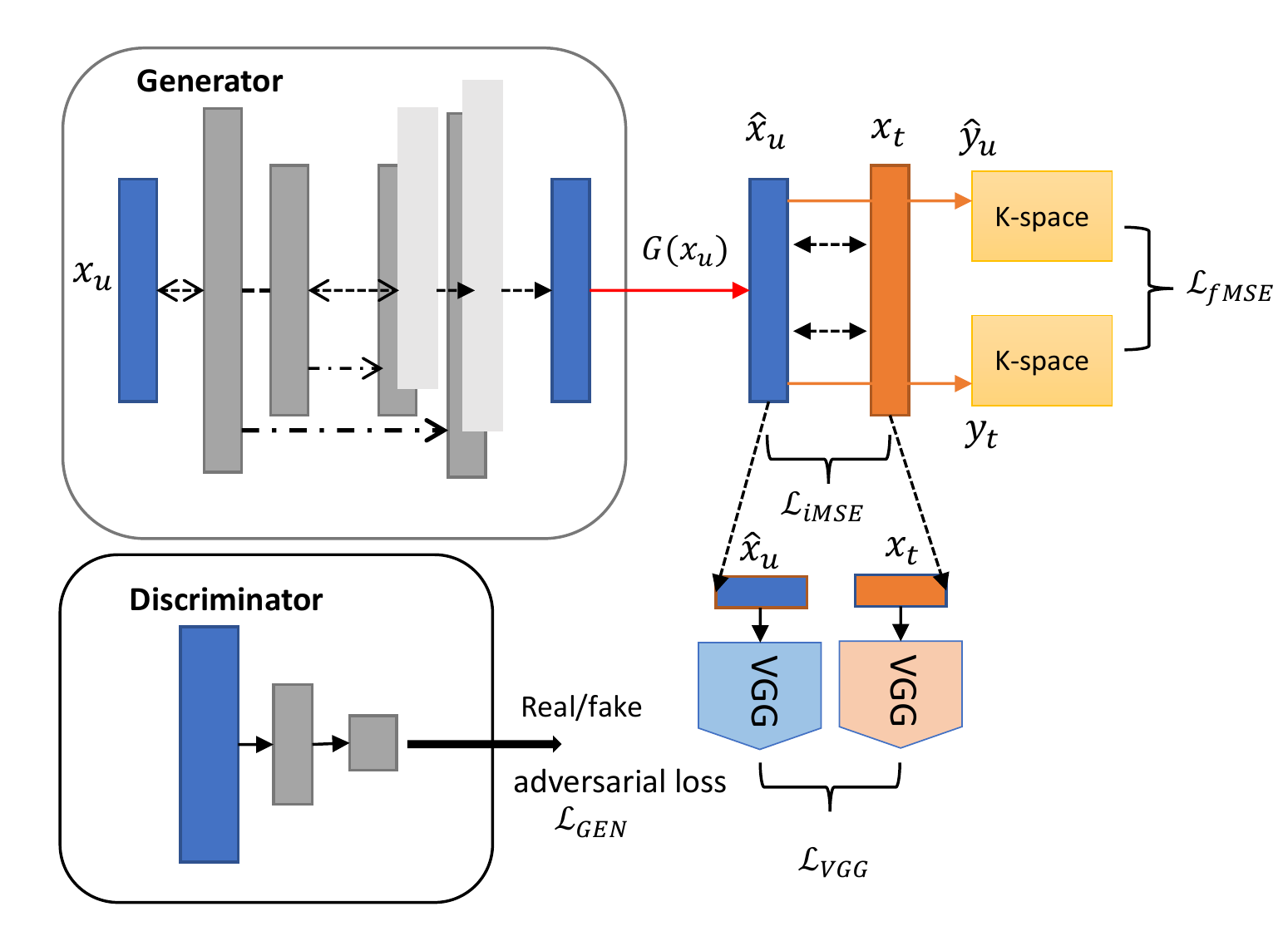}
  \caption{DAGAN: It consists of a generator and discriminator with four loss functions, including loss from pixels and frequency domains, loss from VGG, and adversarial loss. }
\end{figure}

In general, the DAGAN generator is based on the U-Net architecture and consists of 8 convolutional layers for down-sampling and eight corresponding convolutional layers for up-sampling. Each convolutional layer is followed by batch normalization and leaky ReLU layers, and there are skip connections between different levels of the encoder-decoder pairs. The discriminator performs a binary classification task between the generated de-aliased reconstruction image and the ground-truth image. The loss function of DAGAN is based on the GAN training process and can be expressed as:

\begin{align}
    \min_{G}\max_{D}L(G,D) = \mathbb{E}_{x_t\sim p_{train}(x_t)}(\log D(x_t)) \\ \nonumber
    + \mathbb{E}_{x_u\sim p_g(x_u)}(-\log(D(G(x_u)))
\end{align}

Where $x_t$ represents the ground-truth images from fully-sampled k-space and $x_u$ represents the zero-filled image. The generator $G$ produces the corresponding de-aliased reconstruction $G(x_u)=\hat{x}_u$, which is then evaluated by the discriminator $D$. A combined loss was designed for the generator's training to improve the visual quality. This loss includes a pixel-wise mean square error (iMSE) loss, a frequency domain mean square error (fMSE) loss, and a perceptual VGG loss. The dual-MSE loss accounts for both the pixel-wise and frequency domain data, providing effective constraints, while the VGG loss enhances image quality. The two MSE losses for the image domain and frequency domain can be expressed as: 

\begin{equation}
\begin{aligned}
& \min _{G} \mathcal{L}_{\mathrm{iMSE}}\left(G\right)=\frac{1}{2}\left\|\mathrm{x}_t-\hat{\mathrm{x}}_u\right\|_2^2 \\
& \min _{G} \mathcal{L}_{\mathrm{fMSE}}\left(G\right)=\frac{1}{2}\left\|\mathrm{y}_t-\hat{\mathrm{y}}_u\right\|_2^2
\end{aligned}
\end{equation}
Where $\mathrm{\hat{y}_u}$ and $\mathrm{y_t}$ are the corresponding frequency domain data of $\mathrm{\hat{x}_u}$ and $\mathrm{x_t}$ after the Fourier transform. The VGG perceptual loss is represented by:
\begin{equation}
\min _{G} \mathcal{L}_{\mathrm{VGG}}\left(G\right)=\frac{1}{2}\left\|\mathrm{vgg}\left(\mathrm{x}_t\right)-\mathrm{vgg}\left(\hat{\mathrm{x}}_u\right)\right\|_2^2 .
\end{equation}
So, the total combined loss can be written by:
\begin{equation}
\mathcal{L}_{\text {TOTAL }}=\alpha \mathcal{L}_{\mathrm{iMSE}}+\beta \mathcal{L}_{\mathrm{fMSE}}+\gamma \mathcal{L}_{\mathrm{VGG}}+\mathcal{L}_{\mathrm{GEN}} .
\end{equation}
The DAGAN \cite{DGANforCS} loss function for its generator consists of a combination of adversarial loss, pixel-wise mean square error (iMSE) loss, frequency domain mean square error (fMSE) loss and a perceptual VGG loss \cite{VGG}. This combination of losses allows DAGAN to overcome smooth reconstruction difficulties and gain more anatomical information than traditional CNN-based MRI reconstruction algorithms. The adversarial loss is denoted as $\mathcal{L}_{\mathrm{GEN}}$, and $\alpha, \beta,$ and $\gamma$ are hyper-parameters that control the weight of each loss. The dual-MSE and VGG loss have been widely adopted in many subsequent works \cite{jiang2019accelerating, yuan2020sara, lv2021pic, armanious2020medgan}.

\subsection{Recon/RefineGAN}
\begin{figure*}[ht]
\centering
  \includegraphics[scale=0.25]{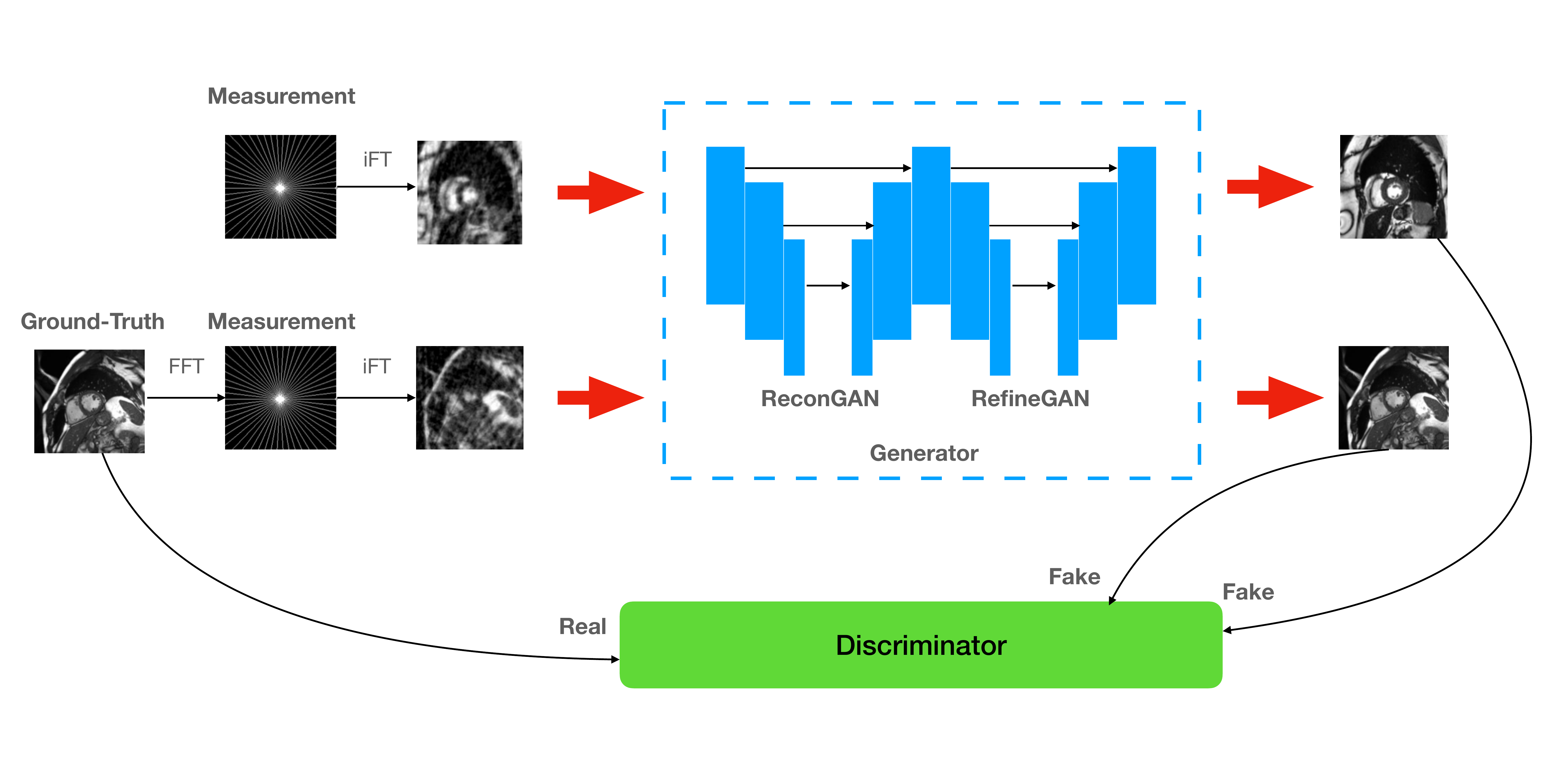}
  \caption{Recon/RefineGAN: The general architecture of Recon/RefineGAN. }
\end{figure*}

Quan et al. \cite{cyclicloss} developed a highly regarded GAN-based MRI reconstruction framework, which blends a deep convolutional auto-encoder with the GAN architecture. This framework enables the utilization of deeper generator and discriminator networks to achieve improved performance. The framework is further enhanced by incorporating cyclic data consistency loss, which ensures faithful interpolation of under-sampled k-space data. Figure 6 shows the details of the Recon/RefineGAN framework. 

The generator $G$ consists of two parts: ReconGAN and RefineGAN, each with an auto-encoder architecture with four encoder and decoder layers. Like DAGAN, Recon/RefineGAN proposed other loss constraints besides a general adversarial loss. They designed a cyclic data consistency loss $L_{cyc}$, which is a combination of under-sampled frequency loss $L_{f}$ and fully reconstructed image loss $L_{img}$ in a cyclic
fashion. The cyclic data consistency loss can be represented by:
\begin{equation}
\begin{aligned}
L_{c y c}(G) & =L_{f }(G)+L_{\text {img }}(G) \\
& =\mathbf{d}(m[i], \bar{m}[i])+\mathbf{d}(s[j], \bar{s}[j])
\end{aligned}
\end{equation}
Where $m[i]$ is the observed under-sampled data from the frequency domain, $\bar{m}[i]$ represents the data in the frequency domain after the $m[i]$ is taken into generator $G$ to get a reconstructed image, then apply under-sampled operator $RF$ (Fourier transform and sampling mask) on it. Similarly, $s[j]$ is the fully reconstructed image; if we use under-sampling operation $RF$ on $s[j]$, then we get $m[j]$ in the frequency domain that can be transformed into zero-filled MR image $s_0[j]$, the $\bar{s}[j]$ represents the image after $s_0[j]$ is taken into generator $G$. $\mathbf{d}$ represents distance metrics, such as mean-square-error
(MSE), and mean-absolute-error (MAE). 

\section{Multi-contrast MRI Reconstruction}

\begin{figure*}[ht]
\centering
  \includegraphics[scale=0.25]{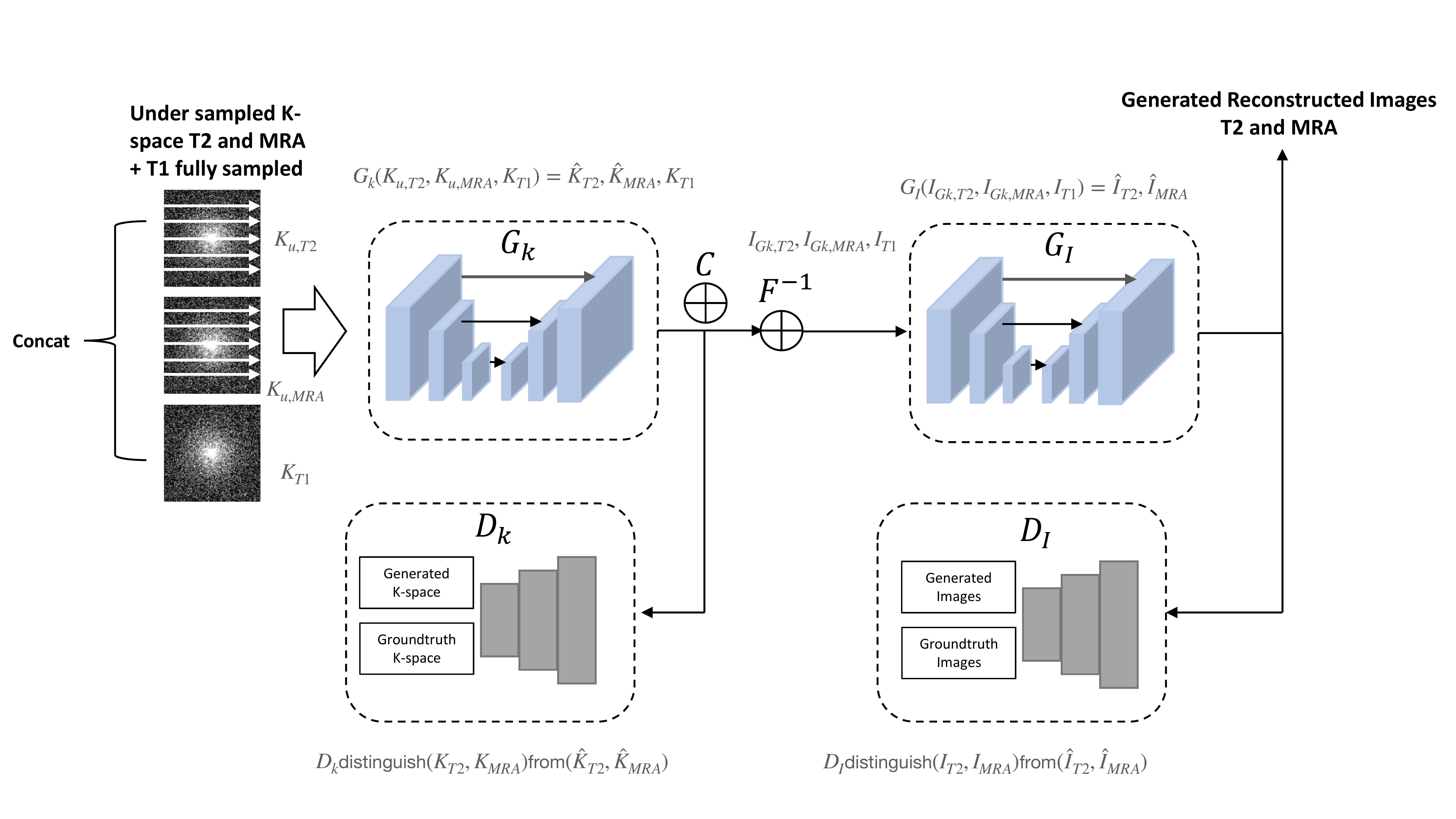}
  \caption{DDGAN: A dual-domain paired GAN networks that consider both image and k-space data}
\end{figure*}
In recent years, the development of various advanced GAN models has led to a surge in research on multi-contrast MRI reconstruction and synthesis. Medical image synthesis has been utilized to facilitate the conversion from one imaging modality to another in clinical settings, with the goal of reducing the amount of radiation exposure for patients. For example, mapping an MR image to a CT image can be considered cross-domain medical image synthesis \cite{wolterink2017deep, wolterink2017mr, hiasa2018cross}. This section will first provide an overview of general medical image synthesis and reconstruction problems and then delve into multiple recent GAN-based architectures that have shown excellent performance in MR image reconstructions.

Multi-contrast MRI is a diagnostic and therapeutic imaging technique that uses multiple imaging sequences within a single MRI scan to produce images with varying contrasts, providing a more in-depth understanding of the body's structure and pathological changes. The imaging sequences, such as T1- and T2-weighted, proton-density-weighted, and contrast-enhanced MRI, each offer distinct information about the scanned tissue, such as the distinction between gray and white matter tissues in T1-weighted brain imaging and fluid and cortical tissue in T2-weighted brain imaging. Magnetic resonance angiography (MRA) is another crucial MR contrast that is used for imaging and assessing vascular anatomy and associated diseases. Different contrasts of MR images often contain similar information, making it possible to reconstruct between contrasts. This versatility of multi-contrast MRI makes it a powerful tool in the medical field \cite{huang2014fast}. 

\subsection{Related Works}
The first deep learning model for multi-contrast CS-MRI reconstruction with fewer model parameters was proposed by Sun et al. \cite{sun2019deep}. Olut et al. \cite{multicontrast1} developed a GAN-based model with a new loss term that accurately captures the directional features of vascular structures and can reconstruct MRA images from T1- and T2-weighted MRI images. Song et al. \cite{song2019coupled} utilized a coupled dictionary learning method to reconstruct T1- and T2-weighted MR images simultaneously. Wei et al. \cite{UndersampledMulti-contrast} created a dual-domain GAN for multi-contrast MRI reconstruction using two pairs of generators and discriminators in both the image and frequency (k-space) domains. Dar et al. \cite{dar2019image} proposed a CGAN-based \cite{cGAN} generative model, which improved the quality and versatility of multi-contrast MRI exams compared to traditional nonlinear regression and deterministic neural networks. It successfully extends the natural image synthesis task into the medical imaging domain. Dar et al. \cite{dar2020prior} developed a new method that builds upon their previous work \cite{dar2019image} and addresses the limitations of pure learning-based reconstruction or synthesis by incorporating three important priors. Yurt et al. \cite{yurt2021mustgan} created a multi-stream GAN architecture for multi-contrast MRI synthesis called MustGAN, which built the adaptive fusion of unique features in multiple one-to-one streams and shared features in a many-to-one stream in order to learn more latent representations and improve the synthesis ability. Do et al. \cite{do2020reconstruction} proposed a dual-GAN-based network to jointly reconstruct T1- and T2-weighted MR images, with the optimization of sampling patterns and training processes verified. 

\subsection{cGAN for medical image synthesis}

In Section II, we introduced the concept of conditional Generative Adversarial Network (cGAN). This technique is commonly used for image-to-image translation and synthesis tasks, such as converting grayscale images to color images and generating full images from partial images. In medical image synthesis, one of the most significant applications is the synthesis of T1- and T2-weighted MR images. Dar et al. \cite{dar2019image} proposed a cGAN-based pipeline for the mutual translation of T1- and T2-weighted MR images. Instead of using the original adversarial loss function defined in equation (4), they utilized a squared adversarial loss function in their approach.

\begin{equation}
\begin{aligned}
L_{\text {c} G A N}(D, G)=-E_{x}\left[(D(x \mid y)-1)^2\right] \\
-E_{z}\left[D(G(z \mid y))^2\right],
\end{aligned}
\end{equation}
Where $x$ represents the ground-truth image, $y$ represents the source image (T1- or T2-weighted) used as a condition, $z$ represents the random noise, and $\hat{x}$ represents the reconstructed image. The cGAN has two branches of GAN pipelines that can translate between T1- and T2-weighted images.

\subsection{Double-Domain GAN}
In this section, we introduce a novel Double-Domain GAN (DDGAN) framework purposed by \textit{Wei} et al. \cite{UndersampledMulti-contrast}, which can reconstruct under-sampled T2-weight and MRA images from fully sampled T1-weighted images in a supervised manner. GAN has been applied in cross-modality image synthesis in recent years. In medical imaging, GAN-based image synthesis models can translate MRI, CT, and PET mutually. DDGAN is inspired by the possible combination of multi-contrast MRI reconstruction with synthesis. Specifically, DDGAN uses fully-sampled T1 k-space data as input with under-sampled k-space data from T2-weighted and MRA, which can share mutual information that improve the reconstruction quality of T2-weighted and MRA images. Moreover, DDGAN was created using a dual-domain GAN architecture, which consists of k-space domain GAN and image domain GAN. The k-space generator (K-generator) encapsulates the data using max-pooling and retains larger feature maps, which are more appropriate for data completion. A classical deep encoding network (U-net) is implemented for the image generator (I-generator) to extract multi-scale and dimensional features in image space for image denoising and enhancement \cite{hyun2018deep, sohail2019unpaired}. Together, these two GANs improved the efficacy of multi-contrast MRI reconstruction.

\textbf{General definition:} To illustrate it clearly, we can use the function to show the target for DDGAN:
\begin{equation}
     \argminD_{\mathbf{\theta}}L\{\textbf{DDGAN}(k_{u,T2},k_{u,MRA},k_{T1} | \theta), I_{T2}, I_{MRA}\}
\end{equation}
Where $L(\cdot)$ represents the loss function, $k_{u, T2}$ and $k_{u, MRA}$ represent under-sampled k-space data of T2-weight and MRA images, while $k_{T1}$ is the fully sampled k-space data of T1-weight images, $I_{T2}$ and $I_{MRA}$ are fully sampled T2-weighted and MRA images in the image domain. The modal \textbf{DDGAN} can generate reconstructed T2-weighted and MRA images which should be indistinguishable from $I_{T2}$ and $I_{MRA}$ from the image space discriminator. In general, the pipeline of DDGAN consists of two modules: the data completion module and the image enhancement module. Figure 7 shows details of DDGAN architecture.

\textbf{Data completion module: }In this module, a specific GAN with the generator and discriminator in the k-space domain are represented by $G_k$ and $D_k$, respectively. Through training, $G_k$ is able to learn a mapping from under-sampled to fully-sampled k-space data, which is then distinguished from ground truth by $D_k$. The input of $G_k$ includes the concatenation of fully-sampled k-space data in the T1-weighted image, under-sampled k-space data from the T2-weighted image, and MRA image. The output of $G_k$ includes a concatenation of fully-sampled k-space T1-weighted image, reconstructed k-space T2-weighted image, and MRA image, the $D_k$ should distinguish the $D_k$ output concatenation image from the ground-truth image (fully-sampled k-space T1-weighted image, T2-weighted image, and MRA image). $C$ is a data consistency operation that can replace the generated k-spaces with the data actually sampled. For GAN in the k-space domain, it provides a pixel-wised $L_2$-loss between generated k-space pairs and standard k-space pairs. The $L_2$-loss function is expressed by:
\begin{align}
    \argminD_{G_k}L_2 = \parallel C \odot G_k(k_{u,T2} \oplus k_{u,MRA} \oplus k_{T1}) \\ \nonumber
    - (k_{T2} \oplus k_{MRA} \oplus k_{}T1) \parallel_2^2
\end{align}
Where, $C \odot$ is the data consistency operation, $\oplus$ is concatenation operation, $\parallel \cdot \parallel_2^2$ is the $L_2$ loss. The  adversarial loss function in the k-space domain can be represented by:
\begin{align}
    \min _{G_k} \max _{D_k} L(G_k,D_k) = \mathbb{E}[\log D_k(k_{\text{T2}},k_{\text{MRA}}, \boldsymbol{k_{\text{T1}}})] \\ \nonumber
    + \mathbb{E}[\log(1-D_k(G_k(k_{\text{u,T2}},k_{\text{u,MRA}}, \boldsymbol{k_{\text{T1}}}))]
\end{align}
Where $\boldsymbol{k_{\text{T1}}}$ is the given fully-sampled T1 image from k-space that can share information with other modalities. 
In general, the data completion module GAN can reconstruct k-space data from the under-sampled one, which can fill up the missing part of the k-space data.

\textbf{Image enhancement module: }In this module, the generator $G_I$ is taken inputs from an inverse Fourier Transform of the concatenation output of $G_k$ in the data completion module, including which means that $G_I$ can learn from fully-reconstructed k-space when the GAN in data complete module is finished training so that the GAN in image domain can focus on image enhancement and de-noising, instead of data translation from k-space. The output of $G_I$ can generate reconstructed T1-weighted MR images, T2-weighted MR images, and MRA images. The loss function in the image enhancement GAN can be represented by:
\begin{align}
    \min _{G_I} \max _{D_I} L(G_I,D_I) = \mathbb{E}[\log D_I(I_{\text{T2}},I_{\text{MRA}}, \boldsymbol{I_{\text{T1}}})] \\ \nonumber
    + \mathbb{E}[\log(1-D_I(G_I(I_{\text{Gk,T2}},I_{\text{Gk,MRA}}, \boldsymbol{I_{\text{Gk,T1}}}))]
\end{align}

Experiments and validation of DDGAN can be divided into three distinct levels. First, it demonstrated the efficacy of dual-GAN architecture by comparing the performance of the k-space generator to that of dual-GAN. (k-space and image-space GAN). Second, it indicates that the completely sampled T1-weighted image is useful for reconstructing T2-weighted and MRA images, which is superior to reconstructing T2-weighted and MRA images using their k-space data alone. Comparing DDGAN to other GAN architectures, such as Pix2Pix, and CS-based MRI reconstruction algorithms revealed that DDGAN outperformed all baselines. In Section VI, experimental details will be provided. 

\section{Experiments}

In this section, we provide an overview of key evaluation metrics for I2I translation and other generative tasks. In addition, we present several essential evaluation metrics that could be used for MRI reconstruction. We identified significant I2I translation experiments, such as Pix2Pix, to demonstrate the benefits and drawbacks of its experimental outcomes and how we can enhance our research based on them. In addition to comparing I2I translation in natural imaging, we select a recent multi-contrast MRI reconstruction pipeline, DDGAN, to demonstrate its experimental results that may have a correlation with natural imaging, as well as to identify its potential advantages and disadvantages in order to direct future research.

\subsection{Evaluation Metrics}

For (I2I) translation tasks, several evaluation metrics are commonly used to measure the quality and similarity between the generated and ground truth images. Some examples of these metrics include:

\textbf{Mean Absolute Error (MAE)}: MAE computes the average absolute difference between the image pixel values of the generated and ground truth images. It can be stated as follows:
\begin{equation}
    \text{MAE} = \frac{1}{n} \sum_{i=1}^n \left| y_i - y_i' \right|
\end{equation}

where $n$ is the number of samples, $y_i$ represents the predicted value for the $i^{th}$ sample and $y_i'$ is the actual value. The absolute difference between these values is calculated using the absolute value operator $\left| \cdot \right|$.

\textbf{Mean Squared Error (MSE)}: MSE is the average squared difference between the image pixel values generated and the ground truth. The MSE is formulated as follows: 

\begin{equation}
    \text{MSE} = \frac{1}{n} \sum_{i=1}^n (y_i - y_i')^2
\end{equation}

The MSE represents the average squared difference between predicted and actual values. In contrast to the MAE, the MSE emphasizes larger differences and is sensitive to outliers. However, MSE and MAE are pixel-level metrics that can only determine the pixel-wise distance; they are hard to measure the structural and textural similarity.

\textbf{FCN score \cite{long2015fully}}: is  an evaluation metric used to quantify the performance of a semantic segmentation model. In particular, people use FCN scores for $\text{semantic maps} \leftrightarrow {\text{real photos}}$ tasks in I2I translation. 

\textbf{Structural Similarity Index (SSIM) \cite{wang2004image}}: SSIM takes into account changes in the mean, standard deviation, and covariance of the pixel values to determine the structural similarity between the generated and ground truth images. SSIM can be described as:
\begin{equation}
    \text{SSIM}(x, y) = \frac{(2\mu_x\mu_y + c_1)(2\sigma_{xy} + c_2)}{(\mu_x^2 + \mu_y^2 + c_1)(\sigma_x^2 + \sigma_y^2 + c_2)}
\end{equation}

The SSIM index evaluates the structural similarity between two images, $x$ and $y$, by examining their mean intensity values ($\mu_x$ and $\mu_y$), the variances of their intensity values ($\sigma_x^2$ and $\sigma_y^2$), and the covariance between the intensity values of the two images ($\sigma_{xy}$). To prevent instability in the division, $c_1$ and $c_2$ are utilized as small constants. The SSIM index ranges from -1 to 1, with a score closer to 1 indicating a high degree of structural similarity between the two images.

\textbf{Peak Signal-to-Noise Ratio (PSNR)}: The PSNR is the ratio between the highest possible pixel value and the mean squared difference between the created image and the ground truth image. PSNR is represented by:
\begin{equation}
    \text{PSNR} = 10\log_{10}\left(\frac{\text{MAX}^2}{\text{MSE}}\right)
\end{equation}

Where MAX is the maximum possible pixel value, MSE is the Mean Squared Error between the predicted and ground truth image. To assess image reconstruction quality, the PSNR compares the maximum possible pixel value with the mean squared error between the predicted and ground truth images. The higher the PSNR, the lower the reconstruction error and the higher the quality of the reconstructed image.

\textbf{Fréchet Inception Distance (FID):} In MRI reconstruction, The FID is a widespread metric used to evaluate the quality of generated images compared to real images. It is defined as:
\begin{equation}
    \operatorname{FID}(P_{r}, P_{g})=\left| \mu_{r}-\mu_{g}\right|{2}^{2}+\operatorname{Tr}\left(\Sigma_{r}+\Sigma_{g}-2\left(\Sigma_{r} \Sigma_{g}\right)^{1 / 2}\right)
\end{equation}

Where $P_{r}$ and $P_{g}$ are probability distributions of real and generated images, respectively. $\mu_{r}$ and $\mu_{g}$ are the mean vectors of the real and generated image features, and $\Sigma_{r}$ and $\Sigma_{g}$ are their covariance matrices.

\textbf{LPIPS (Learned Perceptual Image Patch Similarity)} \cite{zhang2018unreasonable}, which assesses the variety of the translated images and correlates well with a human perceptual similarity. It is calculated as the average LPIPS distance between two sets of translation outputs chosen randomly from the same input. A translated outcome that is more realistic and diverse will have a higher LPIPS score. The function of LPIPS can be expressed by:
\begin{equation}
    d(I_1, I_2) = \frac{1}{W\cdot H} \sum_{x=1}^{W} \sum_{y=1}^{H} \left|f(I_1(x,y)) - f(I_2(x,y))\right|_2
\end{equation}
where $d(I_1, I_2)$ is the LPIPS distance between two images $I_1$ and $I_2$, $W$ and $H$ are the width and height of the images, respectively, and $f(x)$ is the deep neural network used to extract features from the images. The term $\left|f(I_1(x,y)) - f(I_2(x,y))\right|_2$ is the Euclidean distance between the feature vectors extracted from the corresponding pixels in the two images. The final result is the average distance over all the pixels in the images.

\subsection{Datasets}

Table 3 shows some important datasets that are used for I2I translation. Table 4 shows some datasets for MRI reconstruction, including T1, T2, PD, and MRA modalities for heart, knees, and brain research.

\begin{table}[]
   
\caption{Important datasets for I2I translation.}
\resizebox{\columnwidth}{!}{
\begin{tabular}{cccccc}
\toprule
Dataset & Year & Total & Classes & Application \\
\toprule
CelebA\cite{liu2015faceattributes} & 2015 & 202,599 & 10177 & Facial attributes \\
\hline RaFD\cite{langner2010presentation} & 2010 & 8040 & 67 & Facial expressions \\
\hline CMP Facades\cite{Tylecek13} & 2013 & 606 & 12 & Façade images \\
\hline Facescrub\cite{ng2014data} & 2014 & 106,863 & 153 & Faces \\
\hline Cityscapes\cite{Cordts2016Cityscapes} & 2016 & 70,000 & 30 & Semantic \\
\hline Helen Face\cite{le2012interactive} & 2012 & 2330 & $-$ & Face Parsing \\
\hline CartoonSet\cite{royer2020xgan} & 2018 & 10,000 & $-$ & Cartoon Faces \\
\hline
\end{tabular}}
\end{table}

\begin{table}[]
   
\caption{Important datasets for MRI reconstructions.}
\resizebox{\columnwidth}{!}{
\begin{tabular}{cccccc}
\toprule
Dataset & Year & Modality & Regions & Application \\
\toprule
FastMRI\cite{zbontar2018fastmri} & 2018 & PD,T1 & Knees and Brain & AI for fast MRI scan \\
\hline IXI & 2007 & T1,T2,MRA,PD,DTI & Brain & brain research \\
\hline MRBrainS\cite{mendrik2015mrbrains} & 2013 & T1,FLAIR & Brain & MICCAI Challenges \\
\hline ACDC\cite{bernard2018deep} & 2018 & MRI & Heart & heart research challenges \\
\hline
\end{tabular}}
\end{table}

\subsection{Experimental results}

\subsubsection{Pix2Pix}

In Pix2Pix, the author attempts to discover which loss function is more significant using contrastive studies. As described in Section III, Pix2Pix features two loss functions: the adversarial loss, which is used to train the whole network, and the $L_1$ loss, which produces less blurred output than the $L_2$ loss. Table 5 shows that different loss functions could significantly impact generated images evaluated by FCN-Score \cite{long2015fully}. The author also compared the U-Net generator to the conventional auto-encoder generator. Figure 8 indicates that U-net performs better at extracting and retaining high-level features such as textural information. Still, the auto-encoder lacks skip connection structures, and less information may transcend the bottleneck layer, resulting in several repeated lines on output images during upsampling procedures (Block A). In addition, U-net and auto-encoder generators cannot perform well only on $L_1$ loss, but U-Net is still better at capturing structural information (block B).

\begin{table}
\centering
\caption{FCN-scores for different losses, evaluated on Cityscapes dataset}
\begin{tabular}{lccc} 

Loss & Per-pixel acc. & Per-class acc. & Class IOU \\
\hline L1 & $0.42$ & $0.15$ & $0.11$ \\
GAN & $0.22$ & $0.05$ & $0.01$ \\
CGAN & $0.57$ & $0.22$ & $0.16$ \\
L1+GAN & $0.64$ & $0.20$ & $0.15$ \\
L1+CGAN & $\mathbf{0 . 6 6}$ & $\mathbf{0 . 2 3}$ & $\mathbf{0 . 1 7}$ \\
\hline Ground truth & $0.80$ & $0.26$ & $0.21$
\end{tabular}
\end{table}

\begin{figure}[ht]
\centering
  \includegraphics[scale=0.2]{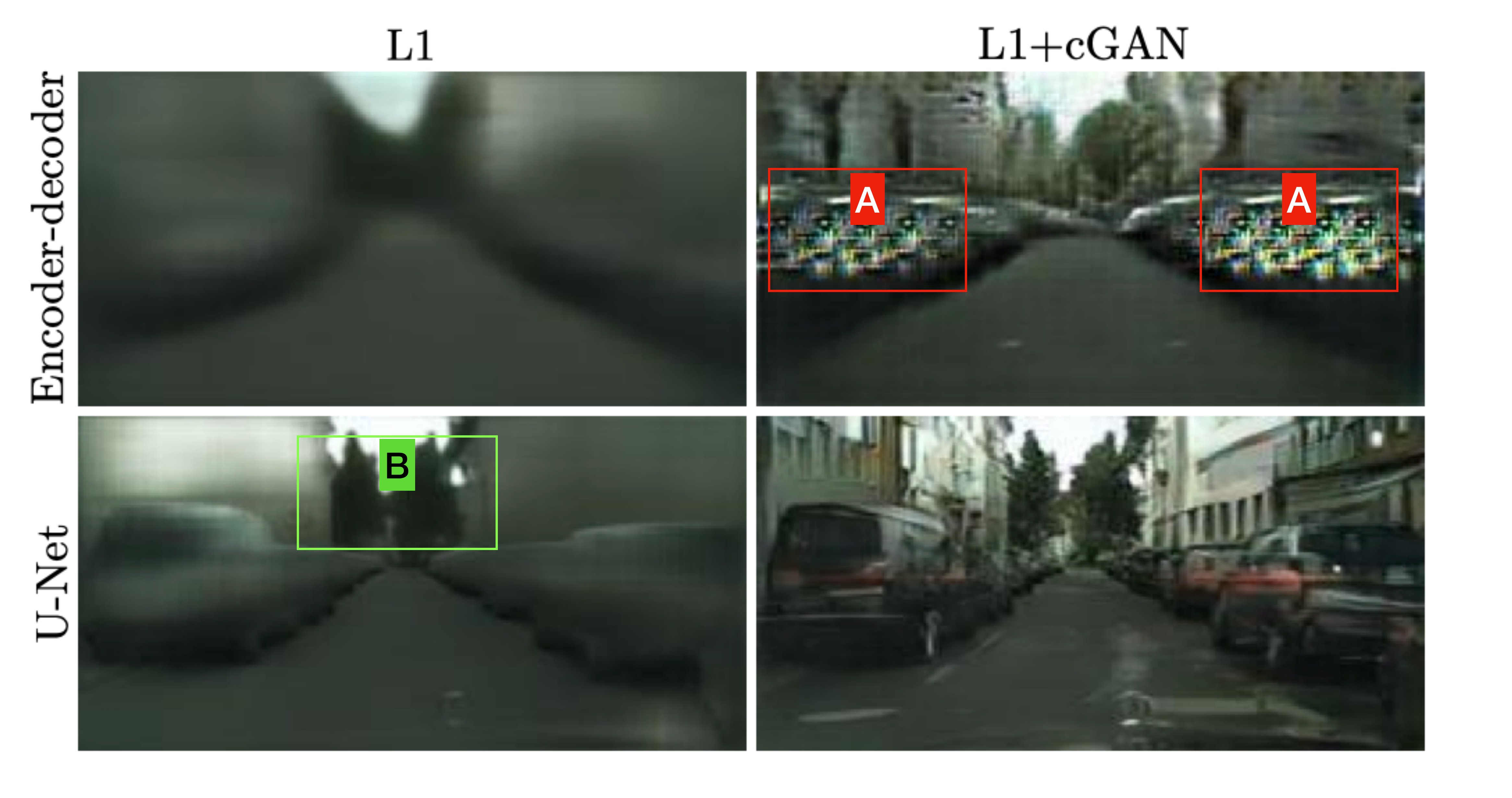}
  \caption{Comparison of U-net generator with auto-encoder under $L_1$ loss and $L_1$+CGAN. \cite{cGANimagetoimage}}
\end{figure}

Furthermore, Pix2Pix has conducted a number of experiments, one of which compared the performance of a PatchGAN-like discriminator using various patch sizes. The results show that a patch size of $70 \times 70$ performed better than a patch size of $286 \times 286$; however, the patch size of $16 \times 16$ performed worse than the patch size of $70 \times 70$. This experiment highlights the potential for decreased performance when using larger patch sizes for the PatchGAN discriminator. From my own perspective, I think PatchGAN has its own contribution to this pipeline so the experimental results to show its performance is important and necessary.  Additionally, the author assessed the perceptual realism of colorization tasks by comparing the performance of different loss functions. The results showed that CGANs could produce compelling colorizations, but a common failure mode is the production of a gray-scale or desaturated result. These experimental results deliver a hint for future research directions that try to fix the problem related to gray-scale images. The author also evaluated the map vs. aerial photograph tasks using the Amazon Mechanical Turk (AMT) method, a perceptual evaluation by human discriminators. The results indicated that aerial photos generated by Pix2Pix could fool 18.9\% of trials, which was significantly better than the baseline. However, maps generated by Pix2Pix only fooled 6.1\% of trials, which could be due to structural errors in the generated maps that are more noticeable when compared to aerial photos. Pix2Pix has also been tested on semantic segmentation tasks, and CGAN with the $L_1$ model has achieved the best performance. In conclusion, when the Pix2Pix paper was published, although Pix2Pix has made a series of experiments, that showed a tremendous number of images as examples, there were fewer evaluation metrics, such as AMT and FCN-based evaluation, which may have defects in determining whether or not the results were satisfactory. This is the primary issue that the Pix2Pix examinations have uncovered.

\subsubsection{DDGAN}

DDGAN \cite{UndersampledMulti-contrast} is a typical multi-contrast MRI reconstruction work, which uses k-space and image domain dual-domain GAN to implement cross-contrast reconstruction for T2-weighted images and MRA by given fully-sampled T1-weighted images. In the experiment part, DDGAN uses the IXI dataset (\url{http://braindevelopment.org/ixi-dataset/}), which contains more than 600 MR images. They created 2D slices from  T1-, T2- weighted, and MRA images and interpolate them into the same dimension and size, which is suitable for the network inputs. Finally, there are 4339 (80\%) slices for training while 1101 (20\%) slices for testing. Figure 9 \cite{UndersampledMulti-contrast} shows examples of image slices and under-sampling masks. 

\begin{figure}[ht]
\centering
  \includegraphics[scale=0.6]{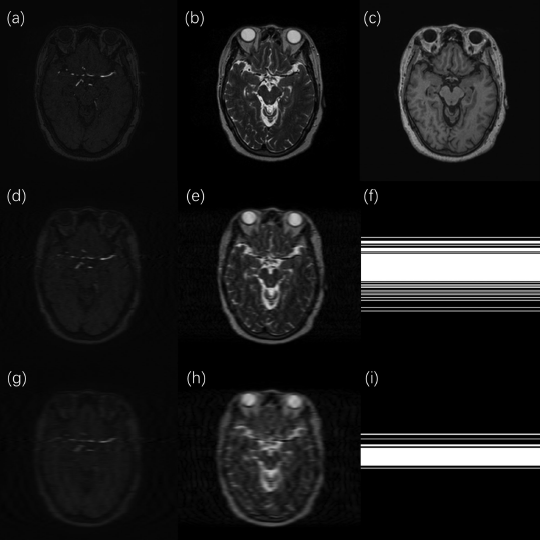}
  \caption{Dataset contains different levels of under-sampling. (a)-(c) are examples of fully-sampled MRA, T2-weighted, and T1-weighted images. (d-e) under-sampled MRA and T2 under (f) mask pattern (rate=4). (g-h) under-sampled MRA and T2 under-sampled rate 8. (i) mask pattern (rate=8). \cite{UndersampledMulti-contrast}}
\end{figure}

The author conducted multiple comparative experiments to demonstrate that DDGAN has exceeded its capacity. First, an experimental result demonstrates that DDGAN outperforms single-domain GAN, including K-space GAN and image-domain GAN, on all three evaluation metrics (PSNR, SSIM, RMSE) and various accelerating rates ($\times{4, 8}$). Second, to demonstrate the effectiveness of multi-input for reconstruction, the author built two models: one that directly reconstructs T2 and MRA images from their under-sampled k-space data, which is called rNet, and another that uses T1 contrast as source information to synthesize T2 and MRA images, which is called sNet. DDGAN performs substantially better than rNet and sNet. These results demonstrate that DDGAN employs completely sampled T1-weighted images as a single input that can share mutual information, which is advantageous when reconstructing other contrast images. Lastly, DDGAN compares its results to those of other models, such as CS-MRI-based traditional methods and the Pix2Pix-based model, which are used as benchmarks to demonstrate DDGAN's success. Figure 10 shows the results comparison of different level baselines and DDGAN. 

\begin{figure}[ht]
\centering
  \includegraphics[scale=0.45]{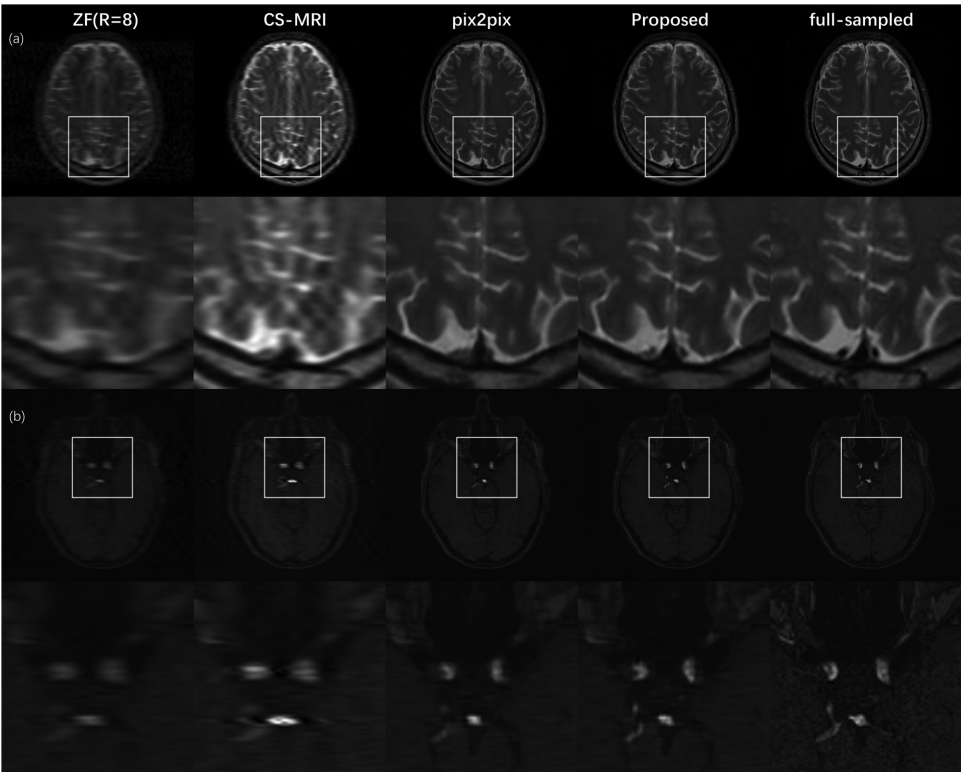}
  \caption{Comparison of zero-filled reconstruction, CS-MRI, Pix2Pix, DDGAN and fully-sampled. \cite{UndersampledMulti-contrast}}
\end{figure}

In conclusion, in my opinion, DDGAN has conducted a series of experiments that demonstrate the efficacy of its concept to combine the fully-sampled T1-weighted image as a successful image source. Furthermore, DDGAN has demonstrated that its performance can outperform some well-known architectures, such as Pix2Pix. When the undersampled mask pattern alters, DDGAN has also demonstrated its reconstruction details. However, DDGAN's experiments are mainly focused on the performance of MRI reconstruction after the mutual information shared by multi-contrast training examples. However, it is hard to show that the model's robustness, portability, and transferability. In the meantime, the GAN-based baselines of DDGAN are confined to Pix2Pix, and it is unknown how DDGAN compares to other pipelines that use multi-contrast MRI as research targets. DDGAN should, for instance, compare itself to Cycle-MedGAN \cite{armanious2019unsupervised}. In addition, DDGAN should attempt to demonstrate its ability to aid in clinical diagnosis by providing examples of its performance in actual diagnostic procedures.

\section{Discussion}

In our previous discussion, we touched upon the various interconnections and associations between natural image-to-image (I2I) translation and its implementation in the area of medical imaging. For instance, certain studies have adapted Pix2Pix for use in the medical field, effectively addressing medical image translation challenges \cite{armanious2020medgan}. Additionally, research utilizing CycleGAN in an unsupervised manner for medical image translation \cite{armanious2019unsupervised} demonstrates that cycle-consistency loss can be employed for image registration during bi-domain medical image translation. With the development of deep learning models in the field of natural imaging, there are an increasing number of works in the medical domain that use Pix2Pix or other well-known benchmarks as a starting point to enhance their work based on the prior state of the art. However, medical imaging has unique characteristics that may hinder the model's performance in the natural image domain. For instance, some medical images have a distinct pattern that reveals information about specific disorders; therefore, our generative model must acquire this pattern effectively and precisely. In this section, we will discuss some significant challenges involving both the natural and medical fields, with the goal of defining a clearer and more distinct boundary that can guide our future research efforts.

\subsection{Reviews in I2I translation}
In the field of natural I2I translation, Pix2Pix has overcome several challenges and problems faced by researchers. Its most significant contribution is integrating separate, special-purpose machinery methods into a single framework that can solve different problems using a single pipeline \cite{zhu2017unpaired}. Pix2Pix uses a U-Net instead of an auto-encoder as the generator's backbone, which allows it to avoid losing structural information and preserve more details and textures. Additionally, Pix2Pix's open-source software allows a wide range of tasks to be demonstrated without needing specific parameter tuning. Pix2Pix has become a successful benchmark in the I2I translation field, as evidenced by its use by various individuals, including artists, for their own purposes. However, Pix2Pix has limitations, such as the need for high-quality paired image data for training, which restricts the model's ability to create diverse outputs. To address this, the same research group as Pix2Pix, Zhu et al., designed CycleGANs for I2I translation without using paired data as training samples, as the generators are trained using only unpaired data from both domains \cite{zhu2017unpaired}. As a result, CycleGAN is a versatile tool for I2I translation when paired data is difficult or impossible to obtain. However, CycleGANs cannot fully bridge the gap between paired and unpaired data models, which means that the performance of models that use paired data is usually better than models using unpaired data. Researchers try to solve this by using weak or semi-supervised data, which could lead to a more effective solution, although labeling data is a time-consuming process. For instance, Park et al. achieved better results in I2I translation using unpaired data through contrastive learning and collaboration with the encoder-decoder, based on InfoNCE loss \cite{oord2018representation, park2020contrastive}.

Furthermore, Pix2Pix and its improved frameworks have limitations in generating multiple outputs or across multiple domains. To overcome this, Choi et al. introduced StarGAN v1 and v2 \cite{choi2018stargan, choi2020stargan}, effectively addressing the multi-domain I2I translation problem. StarGAN uses a single pair of generators and discriminators to learn all mappings and generate multi-domain images. He et al. proposed the deliberation learning framework for I2I translation, which reviews and refines output images, building upon CycleGAN and StarGAN \cite{he2019deliberation}.

Preserving fine-grained details, such as texture and object boundaries, in the output image is often a challenge in I2I translation, and Pix2Pix also faces this limitation. To address this, HDPix2Pix was introduced, which can generate photo-realistic outputs by using a multi-scale generator network to handle complex structures in the output images \cite{wang2018high}. Park et al. proposed the SPADE method, a conditional normalization method that provides better semantic information preservation than traditional normalization layers \cite{park2019semantic}. Lee et al. created DRIT++, a framework for disentangled representation that allows for preserving different image information and uses cross-cycle consistency loss for cyclic reconstruction training without paired data \cite{lee2020drit++}. Shen et al. \cite{shen2019towards} claim that previous approaches, such as DRIT++ and Pix2PixHD, can handle complex structures in the output photos but may have serious limitations if the target domain images contain rich content and multiple disparate objects. They proposed an instance-aware approach that applies fine-grained local and global styles to the target picture spatially and improves the overall quality of the output through joint training.

In general, the challenges associated with I2I translation include the quality of the translation, the preservation of complex context and image structures, multi-domain I2I translation, and training with unpaired data. Researchers continue to enhance the performance of I2I translation tasks by refining algorithms and architectures, selecting suitable loss functions, integrating with sophisticated statistical approaches, and developing unique training pipelines.

\subsection{Reviews of MRI reconstruction}

Here, we highlight some of the key challenges researchers face in MRI reconstruction and suggest potential solutions based on a literature review. A major challenge is improving the perceptual quality of reconstructed images. Training networks solely on L1/L2 distance enforces overall image or patch similarity but fails to address local features such as anatomical sharpness, leading to blurring and fuzzy output. To address this, DAGAN \cite{yang2017dagan} introduced new loss functions, such as the perceptual loss generated by a pre-trained VGG network, to extract high-level features from MRI images and enhance image quality and anatomical information. Chen et al. \cite{chen2020data} proposed a novel encoder-decoder architecture that aims to increase reconstruction quality by preserving local features and sharpening anatomy by confining the GAN to follow the data distribution in the K-space, reducing the risk of artificial details. Li et al. \cite{li2021high} designed an edge-enhanced dual discriminator GAN that improves the reconstruction of edge details, preserving structural details and enhancing overall image quality. Higher-quality MRI reconstruction with fewer anti-aliasing artifacts may lead to more effective therapeutic applications.

Most MRI reconstruction algorithms reviewed rely on supervised learning and thus require fully-sampled ground-truth images, which can be time-consuming to acquire. One potential solution is to develop a semi-supervised or unsupervised learning model, although unsupervised learning often leads to subpar reconstruction quality. Bora et al. \cite{bora2018ambientgan} proposed AmbientGAN, a statistical-based method for solving unsupervised image reconstruction using natural imaging datasets, opening up future MRI reconstruction research opportunities. Cole et al. \cite{cole2020unsupervised} proposed a novel GAN method that quickly generates high-quality 3D MR images compared to baselines. When tested on a knee dataset, the gap between the proposed unsupervised GAN and the supervised GAN was relatively small, with a difference of 0.78\% worse PSNR, 4.17\% worse NRMSE, and equal SSIM. Efforts to reconstruct MR images without fully-sampled data may be faster and more accurate than conventional CS-MRI techniques. However, the state-of-the-art performance and results of supervised MRI reconstruction are still maintained.

Some studies have shifted their focus to image synthesis rather than reconstruction. For example, Sohail et al. \cite{sohail2019unpaired} developed a GAN model to produce fake MR images across multiple modalities and contrasts. These fake images are newly generated signals rather than reconstructed ones, and thus the focus is solely on image synthesis and not reconstruction in the image domain. The ultimate goal of multi-contrast MRI reconstruction is to use the information from one fully-sampled contrast to reconstruct other MR modalities from the given sub-sampled k-space data. Similarly, \cite{dar2019image} proposed a T1- and T2-weighted MR images mutual synthesis pipeline using only the image domain. Wei et al. \cite{UndersampledMulti-contrast} proposed a dual-domain GAN (DDGAN) that uses both fully sampled and subsampled k-space for optimal multi-contrast MRI reconstruction. Multiple contributions have been made by DDGAN, including the reconstruction of T2-weighted and MRA images using fully-sampled T1 and sub-sampled T2 and MRA k-space data. This demonstrates that thoroughly sampled T1-weighted k-space data can provide mutual information that facilitates the enhanced reconstruction of other modalities. Moreover, three-level experimental results indicate that DDGAN obtains the highest score across all evaluation metrics, demonstrating superior performance in recovering intricate structures and minimizing aliasing artifacts for accelerated multi-contrast MRI. But DDGAN has some problems, even though it is one of the newest and most advanced architectures for multi-contrast MRI reconstruction. The author has provided evaluation metrics that emphasize performance and precision, but they lack training performance. Was this DDGAN simple to train? Moreover, this architecture has incorporated fully sampled T1-weighted image pairs as a supervised term, which may be limited in terms of the data source. Finally, the author states that the I-generator requires larger feature maps for data completion, so using ViT-based architecture should be better than traditional CNNs because the self-attention layer in ViT makes it possible to embed information globally across the overall image \cite{zhao2023swingan, visiontransformer}.

\section{Future directions}

GAN-based I2I translation and MRI reconstruction have numerous redesign opportunities. Improving the efficacy of I2I translation and MRI reconstruction can generally be accomplished in two significant ways: we can design novel regularization methods and loss functions to enhance the model's portability and robustness, and we can change the architecture of GANs; for example, we can change the CNN-based generator and discriminator to ViTs. Second, we can abandon GAN-based generative models in favor of the diffusion model, which has recently gained popularity.

Currently, GAN models utilize various CNN architectures such as U-Net, ResNet, auto-encoders, etc., primarily focusing on convolutional layers for extracting low and high-level features from input images. However, Zhang et al. \cite{zhang2019self} introduced a self-attention GAN (SAGAN) that incorporates self-attention into the GAN framework. This is crucial in addressing long-range dependencies as it allows the model to consider relationships between all elements in the input sequence, regardless of their position. Recently, the ViT-based GAN architecture has gained popularity, as using self-attention mechanisms and Transformer models \cite{transformer} has improved performance in various computer vision tasks. Lee et al. \cite{lee2021vitgan} took this a step further by using the ViT architecture as the backbone for both the generator and discriminator, paving the way for fully replacing CNN models in GAN frameworks with ViTs. However, the application of ViT-GAN models, which have primarily focused on natural image I2I translation and other generative tasks, has yet to be extended to the medical imaging domain. VTGAN \cite{kamran2021vtgan} has made a step further toward medical image synthesis using ViT-based GAN architecture, it simultaneously performs semi-supervised image synthesis from fundus photographs to fluorescein angiography (FA) for the diagnosis of retinal disease. 

Our review of related works on ViT-GAN models for MRI reconstruction and other medical imaging generative tasks \cite{SVtrans} \cite{vitganfaster} showed improved performance compared to state-of-the-art CNN-based GAN models. Additionally, the recent use of ViT-GAN for self-supervised multi-contrast MRI reconstruction has opened up new research directions. Zhou et al. \cite{Zhou_2023_WACV} proposed a novel self-supervised framework based on the swin-transformer, known as DSFormer, which leverages information sharing between different modalities to reconstruct MR images. This framework represents the most recent and innovative application of ViT-GAN for MRI reconstruction tasks. Moreover, we have lots of other future directions for GAN-based MRI reconstruction. Researchers may continue to develop more advanced GAN architectures that can handle more complex MRI data and provide higher-quality reconstructions. This may include incorporating additional loss functions or designing new training methods for robustness training. 

On the other hand, the GAN-based generative model is not the only one that can be used to solve I2I translation and MRI reconstruction issues. The diffusion model, also known as denoising diffusion probabilistic models (DDPM) \cite{ho2020denoising}, is a novel Markov-chain-based generative model that can be utilized effectively for a variety of generative tasks, including I2I translation, generating high-resolution images and audio signals, etc. The fundamental concept underlying DDPMs is to characterize the noise distribution probabilistically. Given the observed data, the model is trained to understand the probability distribution of the noise. After learning this distribution, the model can be used to generate new images or audio signals by introducing noise and then removing it using the learned noise distribution. Sasaki et al. \cite{sasaki2021unit} purposed a novel generative model called UNIT-DDPM for unpaired I2I translation. It integrated the de-noising Markov Chain Monte Carlo approach to generate high-resolution outputs, which can outperform traditional GAN-based models by evaluation using novel FID metrics. In the medical imaging domain, diffusion-model-based generative methods are widely used for different objectives, one of the most interesting achievements, which is just accepted by CVPR 2023 is that Takagi et al. \cite{takagi2022high} is able to reconstruct a sequence of images that illustrate what the signal includes by applying the latent diffusion model to the visual fMRI signal; in other words, this work is able to materialize the images that the human brain recalled.

In addition, We can also apply interpretable deep learning techniques and models for MRI reconstruction and get a better interpretation and understanding of our models. The interpretable AI (XAI) has been widely used in the medical imaging domain \cite{abeyagunasekera2022lisa}, but has not been successfully used for MRI reconstruction tasks. Our research can still focus on developing real-time MRI reconstruction software to integrate our new clinical algorithms.

\section{Conclusion}

In our survey, we thoroughly examined various aspects of I2I translation in natural imaging and the challenges and premises of MRI reconstruction. Additionally, we explored the potential links between the natural image domain and the medical image domain in the context of generative deep learning techniques based on GANs. Our survey highlighted several key frameworks that have achieved state-of-the-art performance in the field of MRI reconstruction. We also identified emerging trends and potential directions for future research in this field.

\bibliographystyle{abbrv}
\bibliography{refs}

\begin{thebibliography}{100}

\bibitem{abeyagunasekera2022lisa}
S.~H.~P. Abeyagunasekera, Y.~Perera, K.~Chamara, U.~Kaushalya, P.~Sumathipala,
  and O.~Senaweera.
\newblock Lisa: Enhance the explainability of medical images unifying current
  xai techniques.
\newblock In {\em 2022 IEEE 7th International conference for Convergence in
  Technology (I2CT)}, pages 1--9. IEEE, 2022.

\bibitem{almahairi2018augmented}
A.~Almahairi, S.~Rajeshwar, A.~Sordoni, P.~Bachman, and A.~Courville.
\newblock Augmented cyclegan: Learning many-to-many mappings from unpaired
  data.
\newblock In {\em International conference on machine learning}, pages
  195--204. PMLR, 2018.

\bibitem{amodio2019travelgan}
M.~Amodio and S.~Krishnaswamy.
\newblock Travelgan: Image-to-image translation by transformation vector
  learning.
\newblock In {\em Proceedings of the ieee/cvf conference on computer vision and
  pattern recognition}, pages 8983--8992, 2019.

\bibitem{armanious2019unsupervised}
K.~Armanious, C.~Jiang, S.~Abdulatif, T.~K{\"u}stner, S.~Gatidis, and B.~Yang.
\newblock Unsupervised medical image translation using cycle-medgan.
\newblock In {\em 2019 27th European signal processing conference (EUSIPCO)},
  pages 1--5. IEEE, 2019.

\bibitem{armanious2020medgan}
K.~Armanious, C.~Jiang, M.~Fischer, T.~K{\"u}stner, T.~Hepp, K.~Nikolaou,
  S.~Gatidis, and B.~Yang.
\newblock Medgan: Medical image translation using gans.
\newblock {\em Computerized medical imaging and graphics}, 79:101684, 2020.

\bibitem{bernard2018deep}
O.~Bernard, A.~Lalande, C.~Zotti, F.~Cervenansky, X.~Yang, P.-A. Heng,
  I.~Cetin, K.~Lekadir, O.~Camara, M.~A.~G. Ballester, et~al.
\newblock Deep learning techniques for automatic mri cardiac multi-structures
  segmentation and diagnosis: is the problem solved?
\newblock {\em IEEE transactions on medical imaging}, 37(11):2514--2525, 2018.

\bibitem{bora2018ambientgan}
A.~Bora, E.~Price, and A.~G. Dimakis.
\newblock Ambientgan: Generative models from lossy measurements.
\newblock In {\em International conference on learning representations}, 2018.

\bibitem{diffusion1}
C.~Cao, Z.-X. Cui, S.~Liu, D.~Liang, and Y.~Zhu.
\newblock High-frequency space diffusion models for accelerated mri.
\newblock {\em arXiv preprint arXiv:2208.05481}, 2022.

\bibitem{cao2021remix}
J.~Cao, L.~Hou, M.-H. Yang, R.~He, and Z.~Sun.
\newblock Remix: Towards image-to-image translation with limited data.
\newblock In {\em Proceedings of the IEEE/CVF Conference on Computer Vision and
  Pattern Recognition}, pages 15018--15027, 2021.

\bibitem{chen2020data}
S.~Chen, S.~Sun, X.~Huang, D.~Shen, Q.~Wang, and S.~Liao.
\newblock Data-consistency in latent space and online update strategy to guide
  gan for fast mri reconstruction.
\newblock In {\em Machine Learning for Medical Image Reconstruction: Third
  International Workshop, MLMIR 2020, Held in Conjunction with MICCAI 2020,
  Lima, Peru, October 8, 2020, Proceedings 3}, pages 82--90. Springer, 2020.

\bibitem{chen2018attention}
X.~Chen, C.~Xu, X.~Yang, and D.~Tao.
\newblock Attention-gan for object transfiguration in wild images.
\newblock In {\em Proceedings of the European conference on computer vision
  (ECCV)}, pages 164--180, 2018.

\bibitem{chen2021wavelet}
Y.~Chen, D.~Firmin, and G.~Yang.
\newblock Wavelet improved gan for mri reconstruction.
\newblock In {\em Medical Imaging 2021: Physics of Medical Imaging}, volume
  11595, pages 285--295. SPIE, 2021.

\bibitem{choi2018stargan}
Y.~Choi, M.~Choi, M.~Kim, J.-W. Ha, S.~Kim, and J.~Choo.
\newblock Stargan: Unified generative adversarial networks for multi-domain
  image-to-image translation.
\newblock In {\em Proceedings of the IEEE conference on computer vision and
  pattern recognition}, pages 8789--8797, 2018.

\bibitem{choi2020stargan}
Y.~Choi, Y.~Uh, J.~Yoo, and J.-W. Ha.
\newblock Stargan v2: Diverse image synthesis for multiple domains.
\newblock In {\em Proceedings of the IEEE/CVF conference on computer vision and
  pattern recognition}, pages 8188--8197, 2020.

\bibitem{cole2020unsupervised}
E.~K. Cole, J.~M. Pauly, S.~S. Vasanawala, and F.~Ong.
\newblock Unsupervised mri reconstruction with generative adversarial networks.
\newblock {\em arXiv preprint arXiv:2008.13065}, 2020.

\bibitem{Cordts2016Cityscapes}
M.~Cordts, M.~Omran, S.~Ramos, T.~Rehfeld, M.~Enzweiler, R.~Benenson,
  U.~Franke, S.~Roth, and B.~Schiele.
\newblock The cityscapes dataset for semantic urban scene understanding.
\newblock In {\em Proc. of the IEEE Conference on Computer Vision and Pattern
  Recognition (CVPR)}, 2016.

\bibitem{dar2019image}
S.~U. Dar, M.~Yurt, L.~Karacan, A.~Erdem, E.~Erdem, and T.~Cukur.
\newblock Image synthesis in multi-contrast mri with conditional generative
  adversarial networks.
\newblock {\em IEEE transactions on medical imaging}, 38(10):2375--2388, 2019.

\bibitem{dar2020prior}
S.~U. Dar, M.~Yurt, M.~Shahdloo, M.~E. Ild{\i}z, B.~T{\i}naz, and
  T.~{\c{C}}ukur.
\newblock Prior-guided image reconstruction for accelerated multi-contrast mri
  via generative adversarial networks.
\newblock {\em IEEE Journal of Selected Topics in Signal Processing},
  14(6):1072--1087, 2020.

\bibitem{do2020reconstruction}
W.-J. Do, S.~Seo, Y.~Han, J.~C. Ye, S.~H. Choi, and S.-H. Park.
\newblock Reconstruction of multicontrast mr images through deep learning.
\newblock {\em Medical physics}, 47(3):983--997, 2020.

\bibitem{visiontransformer}
A.~Dosovitskiy, L.~Beyer, A.~Kolesnikov, D.~Weissenborn, X.~Zhai,
  T.~Unterthiner, M.~Dehghani, M.~Minderer, G.~Heigold, S.~Gelly, et~al.
\newblock An image is worth 16x16 words: Transformers for image recognition at
  scale.
\newblock {\em arXiv preprint arXiv:2010.11929}, 2020.

\bibitem{emami2020spa}
H.~Emami, M.~M. Aliabadi, M.~Dong, and R.~B. Chinnam.
\newblock Spa-gan: Spatial attention gan for image-to-image translation.
\newblock {\em IEEE Transactions on Multimedia}, 23:391--401, 2020.

\bibitem{gokaslan2018improving}
A.~Gokaslan, V.~Ramanujan, D.~Ritchie, K.~I. Kim, and J.~Tompkin.
\newblock Improving shape deformation in unsupervised image-to-image
  translation.
\newblock In {\em Proceedings of the European Conference on Computer Vision
  (ECCV)}, pages 649--665, 2018.

\bibitem{GAN}
I.~Goodfellow, J.~Pouget-Abadie, M.~Mirza, B.~Xu, D.~Warde-Farley, S.~Ozair,
  A.~Courville, and Y.~Bengio.
\newblock Generative adversarial networks.
\newblock {\em Communications of the ACM}, 63(11):139--144, 2020.

\bibitem{grappa}
M.~A. Griswold, P.~M. Jakob, R.~M. Heidemann, M.~Nittka, V.~Jellus, J.~Wang,
  B.~Kiefer, and A.~Haase.
\newblock Generalized autocalibrating partially parallel acquisitions (grappa).
\newblock {\em Magnetic Resonance in Medicine: An Official Journal of the
  International Society for Magnetic Resonance in Medicine}, 47(6):1202--1210,
  2002.

\bibitem{ResNet}
K.~He, X.~Zhang, S.~Ren, and J.~Sun.
\newblock Deep residual learning for image recognition.
\newblock In {\em Proceedings of the IEEE conference on computer vision and
  pattern recognition}, pages 770--778, 2016.

\bibitem{he2019deliberation}
T.~He, Y.~Xia, J.~Lin, X.~Tan, D.~He, T.~Qin, and Z.~Chen.
\newblock Deliberation learning for image-to-image translation.
\newblock In {\em IJCAI}, pages 2484--2490, 2019.

\bibitem{kspace}
J.~Hennig.
\newblock K-space sampling strategies.
\newblock {\em European radiology}, 9(6):1020--1031, 1999.

\bibitem{hiasa2018cross}
Y.~Hiasa, Y.~Otake, M.~Takao, T.~Matsuoka, K.~Takashima, A.~Carass, J.~L.
  Prince, N.~Sugano, and Y.~Sato.
\newblock Cross-modality image synthesis from unpaired data using cyclegan:
  Effects of gradient consistency loss and training data size.
\newblock In {\em Simulation and Synthesis in Medical Imaging: Third
  International Workshop, SASHIMI 2018, Held in Conjunction with MICCAI 2018,
  Granada, Spain, September 16, 2018, Proceedings 3}, pages 31--41. Springer,
  2018.

\bibitem{ho2020denoising}
J.~Ho, A.~Jain, and P.~Abbeel.
\newblock Denoising diffusion probabilistic models.
\newblock {\em Advances in Neural Information Processing Systems},
  33:6840--6851, 2020.

\bibitem{huang2014fast}
J.~Huang, C.~Chen, and L.~Axel.
\newblock Fast multi-contrast mri reconstruction.
\newblock {\em Magnetic resonance imaging}, 32(10):1344--1352, 2014.

\bibitem{huang2018multimodal}
X.~Huang, M.-Y. Liu, S.~Belongie, and J.~Kautz.
\newblock Multimodal unsupervised image-to-image translation.
\newblock In {\em Proceedings of the European conference on computer vision
  (ECCV)}, pages 172--189, 2018.

\bibitem{hyun2018deep}
C.~M. Hyun, H.~P. Kim, S.~M. Lee, S.~Lee, and J.~K. Seo.
\newblock Deep learning for undersampled mri reconstruction.
\newblock {\em Physics in Medicine \& Biology}, 63(13):135007, 2018.

\bibitem{cGANimagetoimage}
P.~Isola, J.-Y. Zhu, T.~Zhou, and A.~A. Efros.
\newblock Image-to-image translation with conditional adversarial networks.
\newblock In {\em 2017 IEEE Conference on Computer Vision and Pattern
  Recognition (CVPR)}, pages 5967--5976, 2017.

\bibitem{jaspan2015compressed}
O.~N. Jaspan, R.~Fleysher, and M.~L. Lipton.
\newblock Compressed sensing mri: a review of the clinical literature.
\newblock {\em The British journal of radiology}, 88(1056):20150487, 2015.

\bibitem{jiang2019accelerating}
M.~Jiang, Z.~Yuan, X.~Yang, J.~Zhang, Y.~Gong, L.~Xia, and T.~Li.
\newblock Accelerating cs-mri reconstruction with fine-tuning wasserstein
  generative adversarial network.
\newblock {\em IEEE Access}, 7:152347--152357, 2019.

\bibitem{kamran2021vtgan}
S.~A. Kamran, K.~F. Hossain, A.~Tavakkoli, S.~L. Zuckerbrod, and S.~A. Baker.
\newblock Vtgan: Semi-supervised retinal image synthesis and disease prediction
  using vision transformers.
\newblock In {\em Proceedings of the IEEE/CVF International Conference on
  Computer Vision}, pages 3235--3245, 2021.

\bibitem{kim2019u}
J.~Kim, M.~Kim, H.~Kang, and K.~Lee.
\newblock U-gat-it: Unsupervised generative attentional networks with adaptive
  layer-instance normalization for image-to-image translation.
\newblock {\em arXiv preprint arXiv:1907.10830}, 2019.

\bibitem{kim2017learning}
T.~Kim, M.~Cha, H.~Kim, J.~K. Lee, and J.~Kim.
\newblock Learning to discover cross-domain relations with generative
  adversarial networks.
\newblock In {\em International conference on machine learning}, pages
  1857--1865. PMLR, 2017.

\bibitem{SVtrans}
Y.~Korkmaz, M.~Yurt, S.~U.~H. Dar, M.~{\"O}zbey, and T.~Cukur.
\newblock Deep mri reconstruction with generative vision transformers.
\newblock In {\em International Workshop on Machine Learning for Medical Image
  Reconstruction}, pages 54--64. Springer, 2021.

\bibitem{AlexNet}
A.~Krizhevsky, I.~Sutskever, and G.~E. Hinton.
\newblock Imagenet classification with deep convolutional neural networks.
\newblock In F.~Pereira, C.~Burges, L.~Bottou, and K.~Weinberger, editors, {\em
  Advances in Neural Information Processing Systems}, volume~25. Curran
  Associates, Inc., 2012.

\bibitem{langner2010presentation}
O.~Langner, R.~Dotsch, G.~Bijlstra, D.~H. Wigboldus, S.~T. Hawk, and
  A.~Van~Knippenberg.
\newblock Presentation and validation of the radboud faces database.
\newblock {\em Cognition and emotion}, 24(8):1377--1388, 2010.

\bibitem{le2012interactive}
V.~Le, J.~Brandt, Z.~Lin, L.~Bourdev, and T.~S. Huang.
\newblock Interactive facial feature localization.
\newblock In {\em Computer Vision--ECCV 2012: 12th European Conference on
  Computer Vision, Florence, Italy, October 7-13, 2012, Proceedings, Part III
  12}, pages 679--692. Springer, 2012.

\bibitem{lee2020drit++}
H.-Y. Lee, H.-Y. Tseng, Q.~Mao, J.-B. Huang, Y.-D. Lu, M.~Singh, and M.-H.
  Yang.
\newblock Drit++: Diverse image-to-image translation via disentangled
  representations.
\newblock {\em International Journal of Computer Vision}, 128:2402--2417, 2020.

\bibitem{lee2021vitgan}
K.~Lee, H.~Chang, L.~Jiang, H.~Zhang, Z.~Tu, and C.~Liu.
\newblock Vitgan: Training gans with vision transformers.
\newblock {\em arXiv preprint arXiv:2107.04589}, 2021.

\bibitem{li2014medical}
Q.~Li, W.~Cai, X.~Wang, Y.~Zhou, D.~D. Feng, and M.~Chen.
\newblock Medical image classification with convolutional neural network.
\newblock In {\em 2014 13th international conference on control automation
  robotics \& vision (ICARCV)}, pages 844--848. IEEE, 2014.

\bibitem{li2021high}
Y.~Li, J.~Li, F.~Ma, S.~Du, and Y.~Liu.
\newblock High quality and fast compressed sensing mri reconstruction via
  edge-enhanced dual discriminator generative adversarial network.
\newblock {\em Magnetic Resonance Imaging}, 77:124--136, 2021.

\bibitem{SEGAN}
Z.~Li, T.~Zhang, P.~Wan, and D.~Zhang.
\newblock Segan: structure-enhanced generative adversarial network for
  compressed sensing mri reconstruction.
\newblock In {\em Proceedings of the AAAI Conference on Artificial
  Intelligence}, volume~33, pages 1012--1019, 2019.

\bibitem{vitganfaster}
K.~Lin and R.~Heckel.
\newblock Vision transformers enable fast and robust accelerated mri.
\newblock In {\em International Conference on Medical Imaging with Deep
  Learning}, pages 774--795. PMLR, 2022.

\bibitem{liu2017unsupervised}
M.-Y. Liu, T.~Breuel, and J.~Kautz.
\newblock Unsupervised image-to-image translation networks.
\newblock {\em Advances in neural information processing systems}, 30, 2017.

\bibitem{swintransformer}
Z.~Liu, Y.~Lin, Y.~Cao, H.~Hu, Y.~Wei, Z.~Zhang, S.~Lin, and B.~Guo.
\newblock Swin transformer: Hierarchical vision transformer using shifted
  windows.
\newblock In {\em Proceedings of the IEEE/CVF International Conference on
  Computer Vision}, pages 10012--10022, 2021.

\bibitem{liu2015faceattributes}
Z.~Liu, P.~Luo, X.~Wang, and X.~Tang.
\newblock Deep learning face attributes in the wild.
\newblock In {\em Proceedings of International Conference on Computer Vision
  (ICCV)}, December 2015.

\bibitem{long2015fully}
J.~Long, E.~Shelhamer, and T.~Darrell.
\newblock Fully convolutional networks for semantic segmentation.
\newblock In {\em Proceedings of the IEEE conference on computer vision and
  pattern recognition}, pages 3431--3440, 2015.

\bibitem{luan2017deep}
F.~Luan, S.~Paris, E.~Shechtman, and K.~Bala.
\newblock Deep photo style transfer.
\newblock In {\em Proceedings of the IEEE conference on computer vision and
  pattern recognition}, pages 4990--4998, 2017.

\bibitem{compressedsensing}
M.~Lustig, D.~L. Donoho, J.~M. Santos, and J.~M. Pauly.
\newblock Compressed sensing mri.
\newblock {\em IEEE signal processing magazine}, 25(2):72--82, 2008.

\bibitem{lv2021pic}
J.~Lv, C.~Wang, and G.~Yang.
\newblock Pic-gan: a parallel imaging coupled generative adversarial network
  for accelerated multi-channel mri reconstruction.
\newblock {\em Diagnostics}, 11(1):61, 2021.

\bibitem{lv2021gan}
J.~Lv, J.~Zhu, and G.~Yang.
\newblock Which gan? a comparative study of generative adversarial
  network-based fast mri reconstruction.
\newblock {\em Philosophical Transactions of the Royal Society A},
  379(2200):20200203, 2021.

\bibitem{DGANforCS}
M.~Mardani, E.~Gong, J.~Y. Cheng, S.~S. Vasanawala, G.~Zaharchuk, L.~Xing, and
  J.~M. Pauly.
\newblock Deep generative adversarial neural networks for compressive sensing
  mri.
\newblock {\em IEEE Transactions on Medical Imaging}, 38(1):167--179, 2019.

\bibitem{mendrik2015mrbrains}
A.~M. Mendrik, K.~L. Vincken, H.~J. Kuijf, M.~Breeuwer, W.~H. Bouvy,
  J.~De~Bresser, A.~Alansary, M.~De~Bruijne, A.~Carass, A.~El-Baz, et~al.
\newblock Mrbrains challenge: online evaluation framework for brain image
  segmentation in 3t mri scans.
\newblock {\em Computational intelligence and neuroscience}, 2015:1--1, 2015.

\bibitem{cGAN}
M.~Mirza and S.~Osindero.
\newblock Conditional generative adversarial nets.
\newblock {\em arXiv preprint arXiv:1411.1784}, 2014.

\bibitem{murugesan2019recon}
B.~Murugesan, S.~Vijaya~Raghavan, K.~Sarveswaran, K.~Ram, and M.~Sivaprakasam.
\newblock Recon-glgan: A global-local context based generative adversarial
  network for mri reconstruction.
\newblock In {\em Machine Learning for Medical Image Reconstruction: Second
  International Workshop, MLMIR 2019, Held in Conjunction with MICCAI 2019,
  Shenzhen, China, October 17, 2019, Proceedings 2}, pages 3--15. Springer,
  2019.

\bibitem{ng2014data}
H.-W. Ng and S.~Winkler.
\newblock A data-driven approach to cleaning large face datasets.
\newblock In {\em 2014 IEEE international conference on image processing
  (ICIP)}, pages 343--347. IEEE, 2014.

\bibitem{multicontrast1}
S.~Olut, Y.~H. Sahin, U.~Demir, and G.~Unal.
\newblock Generative adversarial training for mra image synthesis using
  multi-contrast mri.
\newblock In {\em International workshop on predictive intelligence in
  medicine}, pages 147--154. Springer, 2018.

\bibitem{oord2018representation}
A.~v.~d. Oord, Y.~Li, and O.~Vinyals.
\newblock Representation learning with contrastive predictive coding.
\newblock {\em arXiv preprint arXiv:1807.03748}, 2018.

\bibitem{distributedlstm}
H.~Palangi, R.~Ward, and L.~Deng.
\newblock Distributed compressive sensing: A deep learning approach.
\newblock {\em IEEE Transactions on Signal Processing}, 64(17):4504--4518,
  2016.

\bibitem{park2020contrastive}
T.~Park, A.~A. Efros, R.~Zhang, and J.-Y. Zhu.
\newblock Contrastive learning for unpaired image-to-image translation.
\newblock In {\em Computer Vision--ECCV 2020: 16th European Conference,
  Glasgow, UK, August 23--28, 2020, Proceedings, Part IX 16}, pages 319--345.
  Springer, 2020.

\bibitem{park2019semantic}
T.~Park, M.-Y. Liu, T.-C. Wang, and J.-Y. Zhu.
\newblock Semantic image synthesis with spatially-adaptive normalization.
\newblock In {\em Proceedings of the IEEE/CVF conference on computer vision and
  pattern recognition}, pages 2337--2346, 2019.

\bibitem{pathak2016context}
D.~Pathak, P.~Krahenbuhl, J.~Donahue, T.~Darrell, and A.~A. Efros.
\newblock Context encoders: Feature learning by inpainting.
\newblock In {\em Proceedings of the IEEE conference on computer vision and
  pattern recognition}, pages 2536--2544, 2016.

\bibitem{sense}
K.~P. Pruessmann, M.~Weiger, M.~B. Scheidegger, and P.~Boesiger.
\newblock Sense: sensitivity encoding for fast mri.
\newblock {\em Magnetic Resonance in Medicine: An Official Journal of the
  International Society for Magnetic Resonance in Medicine}, 42(5):952--962,
  1999.

\bibitem{cyclicloss}
T.~M. Quan, T.~Nguyen-Duc, and W.-K. Jeong.
\newblock Compressed sensing mri reconstruction using a generative adversarial
  network with a cyclic loss.
\newblock {\em IEEE transactions on medical imaging}, 37(6):1488--1497, 2018.

\bibitem{ren2015faster}
S.~Ren, K.~He, R.~Girshick, and J.~Sun.
\newblock Faster r-cnn: Towards real-time object detection with region proposal
  networks.
\newblock {\em Advances in neural information processing systems}, 28, 2015.

\bibitem{U-Net}
O.~Ronneberger, P.~Fischer, and T.~Brox.
\newblock U-net: Convolutional networks for biomedical image segmentation.
\newblock In {\em International Conference on Medical image computing and
  computer-assisted intervention}, pages 234--241. Springer, 2015.

\bibitem{royer2020xgan}
A.~Royer, K.~Bousmalis, S.~Gouws, F.~Bertsch, I.~Mosseri, F.~Cole, and
  K.~Murphy.
\newblock Xgan: Unsupervised image-to-image translation for many-to-many
  mappings.
\newblock In {\em Domain Adaptation for Visual Understanding}, pages 33--49.
  Springer, 2020.

\bibitem{russakovsky2015imagenet}
O.~Russakovsky, J.~Deng, H.~Su, J.~Krause, S.~Satheesh, S.~Ma, Z.~Huang,
  A.~Karpathy, A.~Khosla, M.~Bernstein, et~al.
\newblock Imagenet large scale visual recognition challenge.
\newblock {\em International journal of computer vision}, 115:211--252, 2015.

\bibitem{sasaki2021unit}
H.~Sasaki, C.~G. Willcocks, and T.~P. Breckon.
\newblock Unit-ddpm: Unpaired image translation with denoising diffusion
  probabilistic models.
\newblock {\em arXiv preprint arXiv:2104.05358}, 2021.

\bibitem{cascadecnn}
J.~Schlemper, J.~Caballero, J.~V. Hajnal, A.~Price, and D.~Rueckert.
\newblock A deep cascade of convolutional neural networks for mr image
  reconstruction.
\newblock In {\em International conference on information processing in medical
  imaging}, pages 647--658. Springer, 2017.

\bibitem{shaul2020subsampled}
R.~Shaul, I.~David, O.~Shitrit, and T.~R. Raviv.
\newblock Subsampled brain mri reconstruction by generative adversarial neural
  networks.
\newblock {\em Medical Image Analysis}, 65:101747, 2020.

\bibitem{GANforBrainMRI}
R.~Shaul, I.~David, O.~Shitrit, and T.~R. Raviv.
\newblock Subsampled brain mri reconstruction by generative adversarial neural
  networks.
\newblock {\em Medical Image Analysis}, 65:101747, 2020.

\bibitem{shen2019towards}
Z.~Shen, M.~Huang, J.~Shi, X.~Xue, and T.~S. Huang.
\newblock Towards instance-level image-to-image translation.
\newblock In {\em Proceedings of the IEEE/CVF conference on computer vision and
  pattern recognition}, pages 3683--3692, 2019.

\bibitem{VGG}
K.~Simonyan and A.~Zisserman.
\newblock Very deep convolutional networks for large-scale image recognition.
\newblock {\em arXiv preprint arXiv:1409.1556}, 2014.

\bibitem{sohail2019unpaired}
M.~Sohail, M.~N. Riaz, J.~Wu, C.~Long, and S.~Li.
\newblock Unpaired multi-contrast mr image synthesis using generative
  adversarial networks.
\newblock In {\em Simulation and Synthesis in Medical Imaging: 4th
  International Workshop, SASHIMI 2019, Held in Conjunction with MICCAI 2019,
  Shenzhen, China, October 13, 2019, Proceedings}, pages 22--31. Springer,
  2019.

\bibitem{song2019coupled}
P.~Song, L.~Weizman, J.~F. Mota, Y.~C. Eldar, and M.~R. Rodrigues.
\newblock Coupled dictionary learning for multi-contrast mri reconstruction.
\newblock {\em IEEE transactions on medical imaging}, 39(3):621--633, 2019.

\bibitem{suarez2017infrared}
P.~L. Su{\'a}rez, A.~D. Sappa, and B.~X. Vintimilla.
\newblock Infrared image colorization based on a triplet dcgan architecture.
\newblock In {\em Proceedings of the IEEE Conference on Computer Vision and
  Pattern Recognition Workshops}, pages 18--23, 2017.

\bibitem{sun2019deep}
L.~Sun, Z.~Fan, X.~Fu, Y.~Huang, X.~Ding, and J.~Paisley.
\newblock A deep information sharing network for multi-contrast compressed
  sensing mri reconstruction.
\newblock {\em IEEE Transactions on Image Processing}, 28(12):6141--6153, 2019.

\bibitem{takagi2022high}
Y.~Takagi and S.~Nishimoto.
\newblock High-resolution image reconstruction with latent diffusion models
  from human brain activity.
\newblock {\em bioRxiv}, pages 2022--11, 2022.

\bibitem{tang2019multi}
H.~Tang, D.~Xu, N.~Sebe, Y.~Wang, J.~J. Corso, and Y.~Yan.
\newblock Multi-channel attention selection gan with cascaded semantic guidance
  for cross-view image translation.
\newblock In {\em Proceedings of the IEEE/CVF conference on computer vision and
  pattern recognition}, pages 2417--2426, 2019.

\bibitem{Tylecek13}
R.~Tyle{\v c}ek and R.~{\v S}{\' a}ra.
\newblock Spatial pattern templates for recognition of objects with regular
  structure.
\newblock In {\em Proc. GCPR}, Saarbrucken, Germany, 2013.

\bibitem{ulyanov2016instance}
D.~Ulyanov, A.~Vedaldi, and V.~Lempitsky.
\newblock Instance normalization: The missing ingredient for fast stylization.
\newblock {\em arXiv preprint arXiv:1607.08022}, 2016.

\bibitem{uzunova2020memory}
H.~Uzunova, J.~Ehrhardt, and H.~Handels.
\newblock Memory-efficient gan-based domain translation of high resolution 3d
  medical images.
\newblock {\em Computerized Medical Imaging and Graphics}, 86:101801, 2020.

\bibitem{transformer}
A.~Vaswani, N.~Shazeer, N.~Parmar, J.~Uszkoreit, L.~Jones, A.~N. Gomez, L.~u.
  Kaiser, and I.~Polosukhin.
\newblock Attention is all you need.
\newblock In I.~Guyon, U.~V. Luxburg, S.~Bengio, H.~Wallach, R.~Fergus,
  S.~Vishwanathan, and R.~Garnett, editors, {\em Advances in Neural Information
  Processing Systems}, volume~30, 2017.

\bibitem{CNN4MRI1}
S.~Wang, Z.~Su, L.~Ying, X.~Peng, S.~Zhu, F.~Liang, D.~Feng, and D.~Liang.
\newblock Accelerating magnetic resonance imaging via deep learning.
\newblock In {\em 2016 IEEE 13th international symposium on biomedical imaging
  (ISBI)}, pages 514--517. IEEE, 2016.

\bibitem{wang2018high}
T.-C. Wang, M.-Y. Liu, J.-Y. Zhu, A.~Tao, J.~Kautz, and B.~Catanzaro.
\newblock High-resolution image synthesis and semantic manipulation with
  conditional gans.
\newblock In {\em Proceedings of the IEEE conference on computer vision and
  pattern recognition}, pages 8798--8807, 2018.

\bibitem{wang2004image}
Z.~Wang, A.~C. Bovik, H.~R. Sheikh, and E.~P. Simoncelli.
\newblock Image quality assessment: from error visibility to structural
  similarity.
\newblock {\em IEEE transactions on image processing}, 13(4):600--612, 2004.

\bibitem{UndersampledMulti-contrast}
H.~Wei, Z.~Li, S.~Wang, and R.~Li.
\newblock Undersampled multi-contrast mri reconstruction based on double-domain
  generative adversarial network.
\newblock {\em IEEE Journal of Biomedical and Health Informatics}, 2022.

\bibitem{wolterink2017mr}
J.~M. Wolterink, A.~M. Dinkla, M.~Savenije, P.~R. Seevinck, C.~van~den Berg,
  and I.~I{\v{s}}gum.
\newblock Mr-to-ct synthesis using cycle-consistent generative adversarial
  networks.
\newblock {\em Proc. Neural Inf. Process. Syst.(NIPS)}, 2017.

\bibitem{wolterink2017deep}
J.~M. Wolterink, A.~M. Dinkla, M.~H. Savenije, P.~R. Seevinck, C.~A. van~den
  Berg, and I.~I{\v{s}}gum.
\newblock Deep mr to ct synthesis using unpaired data.
\newblock In {\em Simulation and Synthesis in Medical Imaging: Second
  International Workshop, SASHIMI 2017, Held in Conjunction with MICCAI 2017,
  Qu{\'e}bec City, QC, Canada, September 10, 2017, Proceedings 2}, pages
  14--23. Springer, 2017.

\bibitem{wu2019transgaga}
W.~Wu, K.~Cao, C.~Li, C.~Qian, and C.~C. Loy.
\newblock Transgaga: Geometry-aware unsupervised image-to-image translation.
\newblock In {\em Proceedings of the IEEE/CVF conference on computer vision and
  pattern recognition}, pages 8012--8021, 2019.

\bibitem{xie2022measurement}
Y.~Xie and Q.~Li.
\newblock Measurement-conditioned denoising diffusion probabilistic model for
  under-sampled medical image reconstruction.
\newblock {\em arXiv preprint arXiv:2203.03623}, 2022.

\bibitem{yang2017dagan}
G.~Yang, S.~Yu, H.~Dong, G.~Slabaugh, P.~L. Dragotti, X.~Ye, F.~Liu,
  S.~Arridge, J.~Keegan, Y.~Guo, et~al.
\newblock Dagan: deep de-aliasing generative adversarial networks for fast
  compressed sensing mri reconstruction.
\newblock {\em IEEE transactions on medical imaging}, 37(6):1310--1321, 2017.

\bibitem{yangCNN4MRI}
Y.~Yang, J.~Sun, H.~Li, and Z.~Xu.
\newblock Admm-csnet: A deep learning approach for image compressive sensing.
\newblock {\em IEEE transactions on pattern analysis and machine intelligence},
  42(3):521--538, 2018.

\bibitem{yi2017dualgan}
Z.~Yi, H.~Zhang, P.~Tan, and M.~Gong.
\newblock Dualgan: Unsupervised dual learning for image-to-image translation.
\newblock In {\em Proceedings of the IEEE international conference on computer
  vision}, pages 2849--2857, 2017.

\bibitem{yuan2018unsupervised}
Y.~Yuan, S.~Liu, J.~Zhang, Y.~Zhang, C.~Dong, and L.~Lin.
\newblock Unsupervised image super-resolution using cycle-in-cycle generative
  adversarial networks.
\newblock In {\em Proceedings of the IEEE Conference on Computer Vision and
  Pattern Recognition Workshops}, pages 701--710, 2018.

\bibitem{yuan2020sara}
Z.~Yuan, M.~Jiang, Y.~Wang, B.~Wei, Y.~Li, P.~Wang, W.~Menpes-Smith, Z.~Niu,
  and G.~Yang.
\newblock Sara-gan: Self-attention and relative average discriminator based
  generative adversarial networks for fast compressed sensing mri
  reconstruction.
\newblock {\em Frontiers in Neuroinformatics}, 14:611666, 2020.

\bibitem{yurt2021mustgan}
M.~Yurt, S.~U. Dar, A.~Erdem, E.~Erdem, K.~K. Oguz, and T.~{\c{C}}ukur.
\newblock mustgan: multi-stream generative adversarial networks for mr image
  synthesis.
\newblock {\em Medical image analysis}, 70:101944, 2021.

\bibitem{zbontar2018fastmri}
J.~Zbontar, F.~Knoll, A.~Sriram, T.~Murrell, Z.~Huang, M.~J. Muckley,
  A.~Defazio, R.~Stern, P.~Johnson, M.~Bruno, et~al.
\newblock fastmri: An open dataset and benchmarks for accelerated mri.
\newblock {\em arXiv preprint arXiv:1811.08839}, 2018.

\bibitem{zhang2019self}
H.~Zhang, I.~Goodfellow, D.~Metaxas, and A.~Odena.
\newblock Self-attention generative adversarial networks.
\newblock In {\em International conference on machine learning}, pages
  7354--7363. PMLR, 2019.

\bibitem{zhang2016colorful}
R.~Zhang, P.~Isola, and A.~A. Efros.
\newblock Colorful image colorization.
\newblock In {\em Computer Vision--ECCV 2016: 14th European Conference,
  Amsterdam, The Netherlands, October 11-14, 2016, Proceedings, Part III 14},
  pages 649--666. Springer, 2016.

\bibitem{zhang2018unreasonable}
R.~Zhang, P.~Isola, A.~A. Efros, E.~Shechtman, and O.~Wang.
\newblock The unreasonable effectiveness of deep features as a perceptual
  metric.
\newblock In {\em Proceedings of the IEEE conference on computer vision and
  pattern recognition}, pages 586--595, 2018.

\bibitem{zhang2017real}
R.~Zhang, J.-Y. Zhu, P.~Isola, X.~Geng, A.~S. Lin, T.~Yu, and A.~A. Efros.
\newblock Real-time user-guided image colorization with learned deep priors.
\newblock {\em arXiv preprint arXiv:1705.02999}, 2017.

\bibitem{zhao2023swingan}
X.~Zhao, T.~Yang, B.~Li, and X.~Zhang.
\newblock Swingan: A dual-domain swin transformer-based generative adversarial
  network for mri reconstruction.
\newblock {\em Computers in Biology and Medicine}, 153:106513, 2023.

\bibitem{Zhou_2023_WACV}
B.~Zhou, N.~Dey, J.~Schlemper, S.~S.~M. Salehi, C.~Liu, J.~S. Duncan, and
  M.~Sofka.
\newblock Dsformer: A dual-domain self-supervised transformer for accelerated
  multi-contrast mri reconstruction.
\newblock In {\em Proceedings of the IEEE/CVF Winter Conference on Applications
  of Computer Vision (WACV)}, pages 4966--4975, January 2023.

\bibitem{zhu2017unpaired}
J.-Y. Zhu, T.~Park, P.~Isola, and A.~A. Efros.
\newblock Unpaired image-to-image translation using cycle-consistent
  adversarial networks.
\newblock In {\em Proceedings of the IEEE international conference on computer
  vision}, pages 2223--2232, 2017.

\bibitem{zhu2020gan}
X.~Zhu, L.~Zhang, L.~Zhang, X.~Liu, Y.~Shen, and S.~Zhao.
\newblock Gan-based image super-resolution with a novel quality loss.
\newblock {\em Mathematical Problems in Engineering}, 2020:1--12, 2020.

\end{thebibliography}
\end{document}